\documentstyle[12pt,aps,floats,epsf]{revtex}

\def\Journal#1#2#3#4{{#1} {\bf #2}, #3 (#4)}

\def\NPB{{\em Nucl. Phys.} B}
\def\PLB{{\em Phys. Lett.}  B}
\def\PRL{\em Phys. Rev. Lett.}
\def\PRD{{\em Phys. Rev.} D}

\def\be{\begin{equation}}
\def\ee{\end{equation}}
\def\bea{\begin{eqnarray}}
\def\eea{\end{eqnarray}}

\begin{document}

\begin{flushright}
JLAB-THY-98-49\\
\end{flushright}

\vspace{0.5cm}

\centerline{\bf FACTORIZATION AND HIGH-ENERGY EFFECTIVE ACTION}

\vspace{1cm}
\centerline{IAN  BALITSKY}
\vspace{0.5cm}
\centerline{ Physics Department, Old Dominion University, Norfolk 
VA 23529} 
\centerline{and}
\centerline{Theory Group, Jefferson Lab, Newport News VA 23606}
\centerline{e-mail: balitsky@jlab.org} 

\vspace{1cm}

\centerline{\bf Abstract} 

\vspace{0.4cm}

I demonstrate that the amplitude for 
high-energy scattering can be factorized as a convolution of 
the contributions 
due to fast and slow fields. The fast and slow fields interact
by means of Wilson-line operators -- infinite gauge factors ordered
along the straight line. The resulting factorization formula gives
a starting point for a new approach to the effective action for 
high-energy scattering in QCD.

\newpage

\section{Introduction}

In the leading logarithmic approximation, the high-energy 
scattering in perturbative QCD is determined by the BFKL 
pomeron \cite{bfkl}. 
It is well known that the power behavior of BFKL cross section 
violates 
the Froissart bound.  The BFKL pomeron describes only 
the pre-asymptotic behavior
at not very large energies and in order to 
find the true high-energy asymptotics 
in perturbative QCD we need to unitarize the BFKL pomeron. 
This is a difficult problem which has been in a need of a solution 
for more than 20 years. However, until recently, it was a common
belief that at least at present energies 
(e.g. for small-x deep inelastic scattering in HERA) the 
corrections to BFKL pomeron are small so they can be neglected. 
Contrary to that expectations,  recent calculation of the 
next-to-leading correction
to the BFKL kernel \cite{nlobfkl} shows that this correction is very big.  
It is very likely that further corrections are also large
 which means that
 we must deal with the problem of 
the unitarization of the BFKL pomeron at present energies. 

One of the most popular ideas on solving this problem is to reduce 
the QCD at high energies to some sort of
two-dimensional effective theory which will be simpler than
the original QCD, maybe even to the extent of exact solvability.
Some attempts in this direction were made starting from the work
\cite{verlinde} but the matter is an open issue for the time being.
 In this paper I will describe the new approach to the effective action
 which is based on the factorization in rapidity space 
 for high-energy scattering.
 
 The form of 
factorization is dictated by process kinematics 
(for a review, see \cite{fak}).  A classical example is the
factorization of the structure functions of deep inelastic scattering 
into coefficient functions and parton densities. 
In the case of deep inelastic 
scattering, there are two different
scales of transverse momentum and it is therefore natural to 
factorize the amplitude in the product of contributions of 
hard and soft parts coming from the regions of small and large transverse 
momenta, respectively. On the contrary, in the case of high-energy 
(Regge-type) processes, all the transverse momenta are of the same order of 
magnitude,  but colliding particles strongly differ in rapidity so
 it is natural to factorize in the
rapidity space.

Factorization in rapidity space means that the 
high-energy scattering amplitude can be represented as a convolution of 
contributions due to ``fast" and ``slow" fields. To be precise, we 
choose a certain rapidity $\eta_0$   to be a ``rapidity divide" 
and we call
fields with $\eta>\eta_0$ fast and fields with $\eta<\eta_0$ slow 
where $\eta_0$ lies in the region between spectator 
rapidity $\eta_A$ and target rapidity $\eta_B$. 
(The interpretation of this fields as
fast and slow is literally true only
in the rest frame of the target but we will use this 
terminology for any frame).

To explain what we mean by the factorization in rapidity space let us 
recall the operator expansion for high-energy scattering 
\cite{ing}
where the explicit integration over fast fields gives the coefficient
functions for the Wilson-line operators representing the 
integrals over slow fields. For a 2$\Rightarrow$2 particle
scattering in Regge limit $s\gg m^2$ 
(where $m$ is
a common mass scale for all other momenta in the problem 
$t\sim p_A^2 
 \sim (p'_A)^2\sim p_B^2\sim (p'_B)^2\sim m^2$)
  we have:
\begin{eqnarray}
\lefteqn{A(p_A,p_B\Rightarrow p'_A,p'_B)=}
\label{fla1.1}\\
&\sum\int d^2x_1...d^2x_nC^{i_1...i_n}(x_1,...x_n)
\langle p_B|{\rm Tr}\{U_{i_1}(x_1)...U_{i_n}(x_n)\}|p'_B\rangle
\nonumber
\end{eqnarray}
(As usual, $s=(p_A+p_B)^2$ and $t=(p_A-p'_A)^2$).
Here  $x_i~(i=1,2)$ are the transverse coordinates 
(orthogonal to both $p_{A}$ and $p_{B}$) and 
$U_i(x)=U^{\dagger}(x){i\over g}{\partial\over\partial x_i}U(x)$ where 
the Wilson-line operator $U(x)$ is the 
gauge link ordered along the infinite
straight line corresponding to the ``rapidity divide'' $\eta_0$. Both 
coefficient functions and matrix elements in Eq. (\ref{fla1.1}) depend 
 on the $\eta_0$ but 
this dependence is canceled in the physical amplitude just as the scale 
$\mu$ (separating coefficient functions and matrix elements) disappears 
from the final results for structure functions in case of usual 
factorization.
Typically, we have the factors $\sim (g^2\ln s/m^2-\eta_0)$ coming from
the ``fast" integral and the factors $\sim g^2\eta_0$ coming from
the ``slow" integral so they combine in a usual log factor
$g^2\ln s/m^2$. In the leading log approximation these factors
sum up into the BFKL pomeron (for a review 
see ref. \cite{lobzor}). 

Unlike usual factorization,
the expansion (\ref{fla1.1}) does not have 
the additional meaning of perturbative $vs$ nonperturbative separation 
-- both the coefficient
functions and the matrix elements have perturbative and 
non-perturbative parts. This happens due to the fact that  the 
coupling constant in a
scattering process is is determined by 
the scale of transverse momenta. When we perform 
the usual factorization 
in hard ($k_{\perp}>\mu$) and soft ($k_{\perp}<\mu$) momenta, 
we calculate the 
coefficient functions perturbatively (because 
$\alpha_s(k_{\perp}>\mu)$ is small) whereas
the matrix elements are non-perturbative. Conversely, when we factorize 
the amplitude in rapidity, both fast and slow parts have 
contributions coming from the regions of large and small 
$k_{\perp}$. In this 
sense, coefficient functions and matrix elements enter the expansion 
(\ref{fla1.1}) on equal footing. We could have integrated first over 
slow fields (having the rapidities close to that of $p_B$)
and the expansion would have the form:
\begin{eqnarray}
&A(s,t)=\sum\int d^2x_1...d^2x_nD^{i_1...i_n}(x_1,...x_n)
\langle p_A|{\rm Tr}\{U_{i_1}(x_1)...U_{i_n}(x_n)\}|p'_A\rangle
\label{fla1.2}
\end{eqnarray}
In this case, the coefficient functions $D$ are the results of integration 
over slow fields ant the matrix elements of the $U$ operators contain only the
large rapidities $\eta >\eta_0$. The symmetry between 
Eqs. (1) and (2)
calls for a factorization formula which would have this symmetry between 
slow and fast fields in explicit form. 

I will demonstrate that one can combine the operator expansions 
(\ref{fla1.1}) and (\ref{fla1.2}) in the following way:
\begin{eqnarray}
\lefteqn{A(s,t)=\sum{i^n\over n!}\int d^2x_1...d^2x_n}\label{fla1.3}\\
&\langle p_A|U^{a_1i_1}(x_1)...U^{a_ni_n}(x_n)|p'_A\rangle 
\langle p_B|U^{a_1}_{i_1}(x_1)...U^{a_n}_{i_n}(x_n)|p'_B\rangle
\nonumber
\end{eqnarray}
where $U^a_i\equiv {\mathop{\rm Tr}}(\lambda^aU_i)$ ($\lambda^a$ 
are the Gell-Mann matrices). It is possible 
to rewrite this factorization 
formula in a more visual form if we agree that operators 
$U$ act only on states 
$B$ and $B'$ and introduce the notation $V_i$ for the same operator as 
$U_i$ only acting on the $A$ and $A'$ states:
\begin{eqnarray}
&A(s,t)=\langle p_A|\langle p_B|
\exp\left(i\!\int\! d^2xV^{ai}(x)
U^a_i(x)\right)|p'_A\rangle|p'_B\rangle
\label{fla4}
\end{eqnarray}
\begin{figure}[htb]
\hspace{3cm}
\mbox{
\epsfxsize=7cm
\epsfysize=7cm
\hspace{0cm}
\epsffile{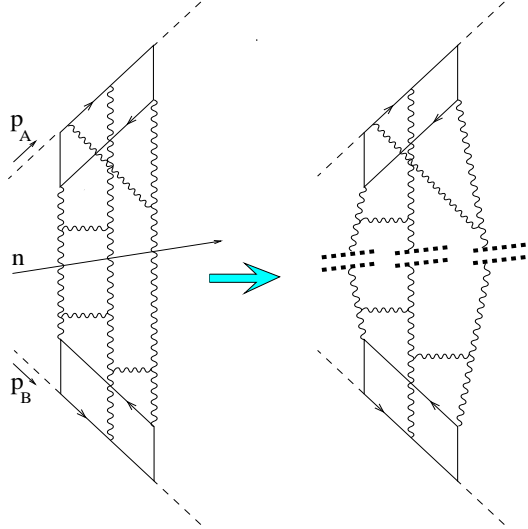}}
\vspace{0.5cm}
{\caption{Structure of the factorization formula. Dashed, solid, and
wavy lines denote photons, quarks, and gluons, respectively. Wilson-line
operators are denoted by dotted lines and the vector $n$ gives the 
direction of
the ``rapidity divide" between fast and slow fields.\label{fig:radish}}}
\end{figure}
In a sense, this formula amounts
to writing the coefficient functions in 
Eq. (\ref{fla1.1}) (or Eq. (\ref{fla1.2})) 
as matrix elements of 
Wilson-line operators. 
Eq. (\ref{fla4}) illustrated in Fig.1 is our main tool for factorizing
in rapidity space. 

\begin{figure}[htb]
\hspace{1cm}
\mbox{
\epsfxsize=12cm
\epsfysize=8cm
\hspace{0cm}
\epsffile{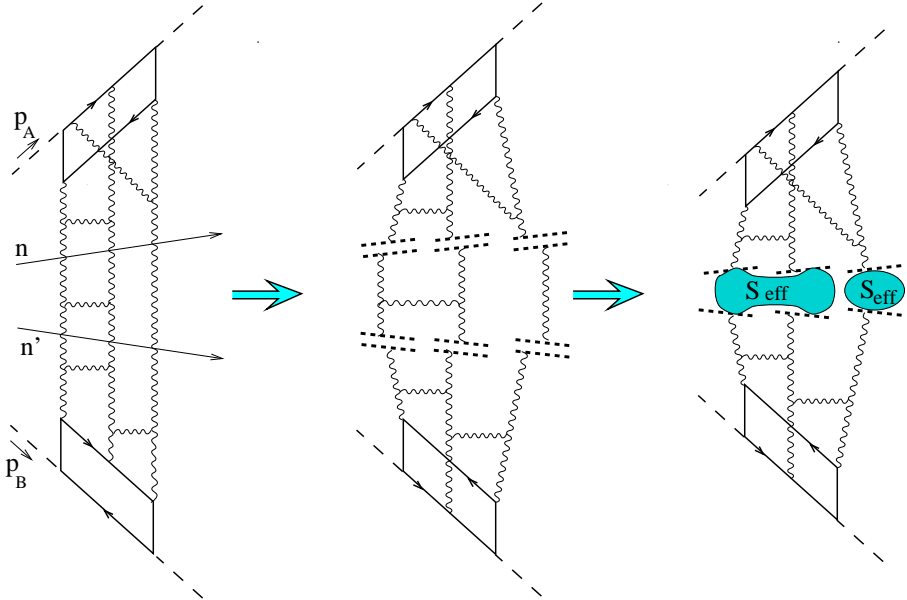}}
\vspace{0.5cm}
{\caption{The effective action for the interval of rapidities 
$\eta_0>\eta>\eta'_0$. The two  vectors
 $n$ and $n'$ correspond to ``rapidity divides" $\eta_0$ and $\eta'_0$ 
bordering our chosen region of rapidities \label{fig2}}}
\end{figure}

In order to define an effective action for a given interval in 
rapidity $\eta_0>\eta>\eta'_0$ we use the master formula (\ref{fla4}) 
two times as illustrated in Fig. \ref{fig2}.
We obtain then
\begin{eqnarray}
&A(s,t)=\langle p_A|\langle p_B|
e^{iS_{\rm eff}(V_i,Y_j)}|p'_A\rangle|p'_B\rangle
\label{fla1.5}
\end{eqnarray}
where the Wilson-line operators $Y_i(x_{\perp})$ have the same form as 
$U_i(x_{\perp})$ but aligned along the $n'$ direction (and act 
only on $B$ and $B'$ states, cf. eq. (\ref{fla4})). In this formula, the 
region
of rapidities greater than $\eta_0$ is represented
operators $V_i$ acting on the  spectator $A$ and $A'$  states, the 
region of rapidities
lower than $\eta'_0$ by the operators $Y_i$ acting on target 
$B$ and $B'$  states, and the region $\eta_0>\eta>\eta'_0$ is integrated
out -all the information about it is contained in the effective action
$S_{\rm eff}(V_i,Y_j)$. As we shall see below, this effective action 
is in general non-local 
(unlike the local interaction term 
$\int dx_{\perp}V_i(x_{\perp})U_i(x_{\perp})$ in the factorization 
formula (\ref{fla4})). Moreover, it contains the factors 
$g^2(\eta_0-\eta_0')$  
which are the usual high-energy logarithms $g^2\ln{s_0\over s'_0}$ 
where the energies $s_0$ and $s'_0$ correspond to rapidities 
$\eta_0$ and $\eta'_0$. If we had a complete expression for 
$S_{\rm eff}(\eta_0,\eta'_0)$ we could take 
$\eta_0=\eta_A$ (rapidity of the spectator particle) and 
$\eta'_0=\eta_B$ (rapidity of the target particle), then all the 
logarithmic dependence on the energy would be included in the effective
action and the resulting matrix elements of the operators $V_i$ between
$A$ states and operators $Y$ between $B$ states will contain no logarithms
(and may me calculated in the first order in perturbation theory 
for a suitable $A$ and $B$ particles such as  virtual photons).
Since multiple rescatterings are taken into account by $e^{iS_{\rm eff}}$ 
automatically the corresponding amplitude must be unitary.
This program is probable not less difficult
than the direct calculation of the many-pomeron exchanges in the 
perturbation theory but for the case of effective-action language
we have some additional powerful methods such as semiclassical approach.

The paper is organized as follows. In Sect. 2 we remind the 
Wilson-line operator language for small-x physics. The 
factorization formula (\ref{fla4}) is derived in Sect. 3
and in Sect. 4 we use it to define formally the high-energy effective action 
for a given interval in rapidity 
(Some of the results of this Sections were reported earlier in the 
letter \cite{prl}). A semiclassical approach to
calculation of this effective action 
is discussed in Sect. 5 and Sect. 6 contains conclusions and outlook.

\section{Operator expansion for high-energy scattering}

Let us now briefly remind how to obtain the operator expansion
(\ref{fla1.1}). For simplicity, 
consider the classical example of high-energy scattering of 
virtual photons with virtualities $\sim -~m^2$. 
\begin{equation}A(s,t)=-i{\mbox{$\langle 0|$}} 
T\{j(p_A)j(p'_A)j(p_B)j(p'_B)\}{\mbox{$|0\rangle $}}.
\label{fla5}
\end{equation}
where $j(p)$ is the Fourier transform of 
electromagnetic current $j_{\mu}(x)$ 
multiplied by some suitable polarization $e^{\mu}(p)$.
At high energies it is convenient to use the Sudakov decomposition:  
\begin{equation}
p^{\mu}~=~\alpha_pp_1^{\mu}+\beta_pp_2^{\mu}+p_{\perp}^{\mu}
\label{suda}
\end{equation}
where
$p_1^{\mu}$ and
$p_2^{\mu}$ are the
light-like vectors close to $p_A$ and 
$p_B$, respectively 
($p_A^{\mu}= p_1^{\mu}-p_2^{\mu}p_A^2/s,~
p_B^{\mu}= p_2^{\mu}-p_1^{\mu}p_B^2/s$).
We want to integrate over the fields with 
$\alpha>\sigma$ where $\sigma$ is
defined in such a way that the corresponding 
rapidity is $\eta_0$. (In explicit form  
$\eta_0=\ln{\sigma\over\tilde{\sigma}}$
where $\tilde{\sigma}\equiv {m^2\over s\sigma}$). 
The result of the integration
will be given by Green functions of the fast particles in 
slow ``external" fields \cite{ing} (see also ref. \cite{larry1}).
 Since the fast particle moves along a 
straight-line classical trajectory 
the propagator is proportional to 
the straight-line ordered gauge factor $U$ \cite{nacht}. For example, when 
$x_{+}>0,~y_{+}<0$ it has the form\cite{ing}:
\begin{eqnarray}
G(x,y)=i\int \! dz\delta
(z_{\ast})
\frac {(\not\!\! x-\not\!\! z)\not\!\!p_2}{2\pi ^{2}(x-z)^{4}}
U(z_{\perp})\frac {\not\!\! z-\not\!\! y}{2\pi
^{2}(z-y)^{4}}
\label{fla6}
\end{eqnarray}
We use the notations 
$z_{\bullet}\equiv z_{\mu}p_1^{\mu}$ and $z_{\ast}\equiv z_{\mu}p_
2^{\mu}$ which 
are essentially identical to the light-front coordinates $z_+=z_*/\sqrt{s},
~z_-=z_{\bullet}/\sqrt{s}$. 
The Wilson-line operator $U$ is defined as
\begin{equation}
U(x_{\perp})=[\infty p_1+x_{\perp}, -\infty p_1+x_{\perp}]
\label{fla7}
\end{equation}
where $[x,y]$ is the straight-line ordered gauge link 
suspended between the points $x$ and $y$:
\begin{eqnarray}
&[x,y]{\stackrel{\rm def}\equiv}
 P\exp\left(ig\int_0^1du (x-y)^{\mu}A_{\mu}(ux+(1-u)y)\right)
\label{fla8}
\end{eqnarray}

The origin of Eq. (\ref{fla6}) is more clear in the rest 
frame of the ``A" photon (see Fig.2). 
\begin{figure}[htb]
\hspace{4cm}
\mbox{
\epsfxsize=5cm
\epsfysize=5cm
\hspace{0cm}
\epsffile{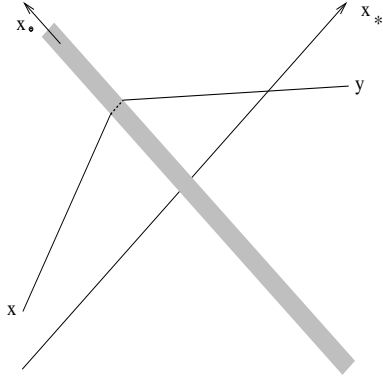}}
\vspace{0.5cm}
{\caption{Quark propagator in a shock-wave background \label{fig3}}}
\end{figure}
Then the quark is slow and the external
fields are approaching this quark at high speed. Due to the Lorentz
contraction, these fields are squeezed in a shock wave located at 
$z_{*}=0$
(in a suitable gauge like the Feynman one).
Therefore,  the propagator (\ref{fla6}) of the quark in this
shock-wave background is a product of three factors which reflect 
(i) free propagation 
from $x$ to the shock wave (ii) instantaneous interaction with the shock 
wave which
is described by the operator $U(z_{\perp})$ 
\footnote{
Because the shock wave is very thin the quark 
has no time to deviate in the transverse direction. Therefore
the quark's trajectory inside the shock wave can be approximated by a 
light-like straight line which means that the interaction of the quark 
with the shock wave will be described by a gauge factor ordered 
along this
segment of a straight line. Since there is no field outside the 
shock-wave "wall" one can formally extend the limits of integration
in a gauge factor to $\pm\infty$ which gives the operator $U$}
, and
(iii) free propagation from 
the point of interaction $z$ to the final destination $y$. 

\begin{figure}[htb]
\hspace{4cm}
\mbox{
\epsfxsize=6cm
\epsfysize=6cm
\hspace{0cm}
\epsffile{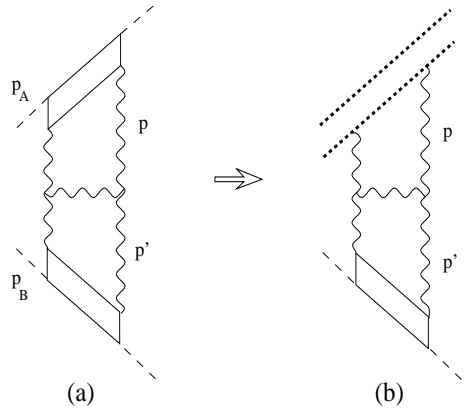}}
\vspace{0cm}
{\caption{A typical Feynman diagram for the $\gamma^*\gamma^*$ scattering 
amplitude (a) and the corresponding two-Wilson-line operator 
(b)\label{fig4}}}
\end{figure}

The propagation of the quark-antiquark pair in the shock-wave 
background is described by the product of two propagators of
Eq. (\ref{fla6}) type which contain two Wilson-line factors 
$U(z)U^{\dagger}(z')$ 
where $z'$ is the point where the antiquark crosses the shock wave. If we
substitute this quark-antiquark propagator in the original expression
for the amplitude (\ref{fla5}) we obtain\cite{ing}:
\begin{eqnarray}
&\int \! d^{4}x  d^{4}z e^{ip_A\cdot x +iq\cdot z} 
   \langle T\{j(x+z)j(z)\}\rangle_{A}
   \simeq \nonumber\\
 &
  \int \frac {d^2p_{\perp}}{4\pi^2}
   I(p_{\perp},q_{\perp}) 
   {\mathop{\rm Tr}}\{ U(p_{\perp})U^{\dagger}(q_{\perp}-p_{\perp}) \} 
\label{fla9}
\end{eqnarray}
where $U(p_{\perp})$ is the Fourier transform of $U(x_{\perp})$ and 
$I(p_{\perp},q_{\perp})$ is the so-called ``impact factor" which is a 
function of $p_{\perp}^2,p_{\perp}\!\cdot\! q_{\perp}$, and photon 
virtuality \cite{mes},\cite{ing}. 
Thus, we have reproduced the leading term in the 
expansion (\ref{fla1.1}). (To recognize it, note that
$U^{\dagger}(x_{\perp})U(y_{\perp})=
P\exp\left\{-ig\int^x_y dz_i U_i(z_{\perp})\right\}$
where the precise form of the path between points
 $x_{\perp}$ and $y_{\perp}$ does not 
matter since this is actually a formula for the 
gauge link in a pure gauge field $U_i(z_{\perp})$). So, in the leading
order in perturbation theory we have calculated the integral over fast 
fields explicitly and reduced the remaining integral over slow fields
to the matrix element of the two-Wilson-line operator, see Fig. \ref{fig4}. 
It is worth noting that
in the next order in perturbation theory we will get 
the contribution to the r.h.s of Eq. (\ref{fla9}) proportional to 
four-Wilson-line operators, in the next to six-line 
operators and so on.
 
 Note that formally we have obtained the operators $U$ ordered along the 
 light-like lines. Matrix elements of such operators contain divergent 
longitudinal integrations which reflect the fact that light-like gauge factor 
corresponds to a quark moving with speed of light (i.e., with infinite 
energy). 
This divergency can be seen already at the one-loop level
if one calculates the contribution to the matrix element of the
two-Wilson-line operator $U(x_{\perp})U^{\dagger}(y_{\perp})$ 
between the "virtual photon states"
shown in Fig. \ref{fig4}.
The reason for this divergency is very simple. 
We have replaced the fast-quark propagator in the "external 
field" (represented here by two gluons 
 coming from the bottom part of the diagram)  by the light-like
Wilson line. In doing so we  have assumed that these two gluons  
are slow, 
$\eta_p\ll\eta_A$.
However, when we calculate the matrix element of the 
$U(x_{\perp})U^{\dagger}(y_{\perp})$ formally the integration
 over the rapidities of the gluon  $\eta_p$ is unbounded so 
 our divergency
 comes from the fast part of the external 
 field which really does not belong there. 
 Indeed, if the rapidity of the gluon 
 $\eta_p$ is of the order of the rapidity of the quark this gluon 
 is a fast one so it 
 will contribute to the coefficient function (in front of the 
 operator constructed from the slow fields) rather than to 
 the matrix element of the operator. 
   
  This is very similar to the case of usual light-cone expansion for the
 deep inelastic scattering (DIS) at moderate x. In that case , we at first
 expand near the light cone (in inverse powers of $Q^2$). The result is
 that the amplitude of DIS is reduced to matrix elements of the 
 light-cone operators which are known as parton densities in the nucleon. 
These matrix elements contain logarithmical divergence in transverse
momenta for the same reason as above - when expanding
around the light cone we assumed that there are no hard quarks and gluons 
inside the proton, but  matrix elements of light-cone operators
contain formally unbounded integrations over $k_{\perp}^2$.  It is well
known how to proceed in this case: we define the renormalized light-cone
operators with the transverse momenta $k_{\perp}^2>\mu^2$ cut off
and expand our T-product of electromagnetic currents in a set these 
renormalized light-cone operators rather than in a set of the original 
unrenormalized
ones (see e.g. \cite{eveq}). After that, the matrix elements of these
operators (parton densities) contain factors $\ln{\mu^2\over m^2}$ and
the corresponding coefficient functions contain $\ln{Q^2\over \mu^2}$.
When we calculate the amplitude we add these factors together so the 
dependence on the 
factorization scale $\mu$ cancels and 
we get the usual DIS logarithmical factors $\ln{Q^2\over m^2}$.
 
\begin{figure}[htb]
\mbox{
\epsfxsize=16cm
\epsfysize=5cm
\hspace{0cm}
\epsffile{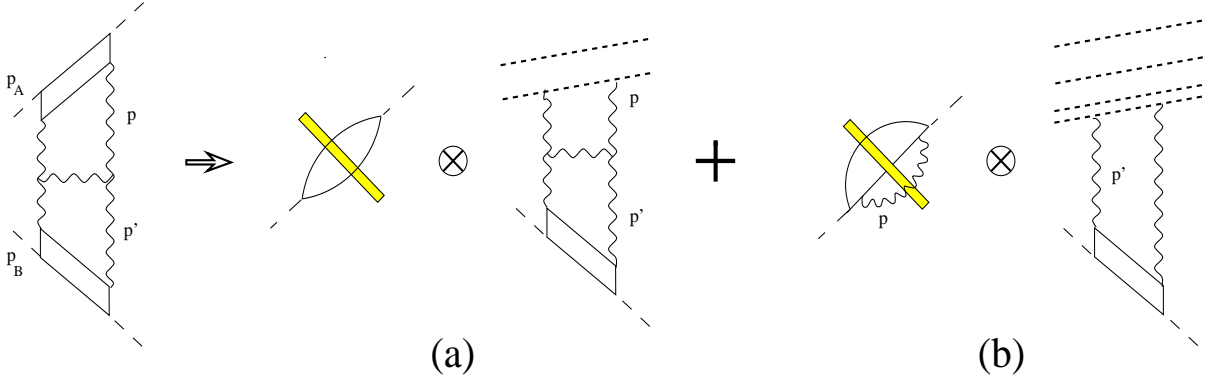}}
\vspace{0.5cm}
{\caption{Decomposition into product of coefficient function
and matrix element of the two-Wilson-line operator for a typical Feynman
diagram. (Double Wilson line  corresponds to fast-moving gluon)\label{figxz}}}
\end{figure}

 Similarly, we need some regularization of the Wilson-line operator which 
 cuts off the fast gluons. 
As demonstrated in \cite{ing},  It can be done
 by changing the slope of the
supporting line as demonstrated in \cite{ing}. If we wish the 
longitudinal integration stop at
$\eta=\eta_0$, we should order our gauge factors $U$ along a 
line parallel
to $n=\sigma p_1+ \tilde{\sigma}p_2$,
then  the coefficient 
functions in front of Wilson-line operators (impact factors) 
will contain logarithms 
$\sim g^2\ln 1/\sigma$.  Similarly to DIS, when we calculate the amplitude,
 we add the terms $\sim g^2\ln 1/\sigma$ coming from
the coefficient functions (see Fig. \ref{figxz}b) to the terms $\sim g^2\ln
{\sigma\over m^2/s}$ coming from matrix elements (see Fig. \ref{figxz}a) 
so
that the dependence on the ``rapidity divide" $\sigma$ cancels and we get the
usual high-energy factors  $g^2\ln {s\over m^2}$ which are responsible for 
BFKL pomeron. 

\section{Factorization formula for high-energy scattering}
In order to understand how this expansion can be generated by the factorization
formula of Eq. (\ref{fla1.3}) type we have to rederive the 
operator expansion in axial gauge $A_{\bullet}=0$ with an additional condition 
$\left.A_{*}\right|_{x_*=-\infty}=0$ (the existence of such a gauge was 
illustrated in \cite{wlup} by an explicit construction). It is important to 
note that with
with power accuracy (up to corrections $\sim \sigma$) our gauge condition may
be replaced by 
$n^{\mu}A_{\mu}=0$. In this gauge the coefficient
functions are given by Feynman diagrams in the external field
\begin{equation}
B_i(x)=U_i(x_{\perp})\Theta(x_*),~~~~~~~~~~~~~~~ B_{\bullet}=B_{*}=0
\label{fla10}
\end{equation} 
which is a gauge rotation of our shock wave (it is easy to see that the 
only nonzero component of the field strength tensor 
$F_{\bullet i}(x)=U_i(x_{\perp})\delta(x_*)$ corresponds to shock wave). 
The Green functions in external field (\ref{fla10}) can be obtained
from a generating functional with a source responsible for this external field. 
Normally, the source for given external field $\bar{{\cal A}}_{\mu}$ is just 
$J_{\nu}=\bar{D}^{\mu} \bar{F}_{\mu\nu}$ so in our case the only non-vanishing 
contribution  is $J_{*}(B)=\bar{D}^i\bar{F}_{i*}$. However, 
we have a problem because the field
which we try to create by this source does not decrease at infinity. To 
illustrate the problem, suppose that we use another light-like gauge
${\cal A}_{*}=0$ for a calculation of the propagators in the external field 
(\ref{fla10}). In this case, the only would-be nonzero contribution
to the source term in the functional integral 
$\bar{D}^i\bar{F}_{i_{\bullet}}{\cal A}_{*}$ vanishes,
 and it looks like 
we do not need a source at all to generate the field $B_{\mu}$!
(This is of course wrong since $B_{\mu}$ is not a classical solution).
What it really means is that the source in this case lies entirely at the 
infinity. Indeed, when we are trying to make an external field 
$\bar{{\cal A}}$ 
in the
functional integral by the source $J_{\mu}$ we need to make a shift
${\cal A}_{\mu}\rightarrow {\cal A}_{\mu}+\bar{{\cal A}}_{\mu}$ 
in the functional integral
\begin{eqnarray}
&\int{\cal D}{\cal A} \exp\left\{iS({\cal A})-i\!\int\! d^4x 
J^a_{\mu}(x){\cal A}^{a\mu}(x)\right\}
\label{fla11}
\end{eqnarray}
after which the linear term $\bar{D}^{\mu} \bar{F}_{\mu\nu}{\cal A}^{\nu}$ 
cancels
with our source term $J_{\mu}{\cal A}^{\mu}$ and the terms quadratic in 
${\cal A}$ 
make the Green functions in the external field $\bar {\cal A}$.
(Note that the classical action $S(\bar{{\cal A}})$ for our external 
field $\bar{{\cal A}}=B$ (\ref{fla10}) vanishes). 
However, in order to reduce the linear
term $\int d^4x\bar{F}^{\mu\nu}\bar{D}_{\mu}{\cal A}_{\nu}$ in the functional 
integral to the form 
$\int d^4x\bar{D}^{\mu} \bar{F}_{\mu\nu}{\cal A}^{\nu}(x)$ we need to make an 
integration by parts, and if the external field does not decrease 
there will be 
additional surface terms at infinity. In our case we are trying to make the
external field $\bar{{\cal A}}=B$ so the linear term which need to be 
canceled by the source is
\begin{eqnarray}
&{2\over s}\int\! dx_{\bullet}dx_{*}d^2x_{\perp} \bar{F}_{i\bullet}
\bar{D}_{*}{\cal A}^{i}=
\left.\int\! dx_{*}d^2x_{\perp} \bar{F}_{i\bullet}
{\cal A}^{i}\right|^{x_{\bullet}=\infty}_{x_{\bullet}=-\infty}
\label{fla12}
\end{eqnarray}
It comes entirely from the boundaries of integration. If we
recall that in our case 
$\bar{F}_{\bullet i}(x)=U_i(x_{\perp})\delta(x_*)$ we can finally rewrite 
the linear term as
\begin{eqnarray}
&\int\! d^2x_{\perp} U_i(x_{\perp})
\{{\cal A}^{i}(-\infty p_2+x_{\perp})-{\cal A}^{i}(\infty p_2+x_{\perp})\}
\label{fla13}
\end{eqnarray}
The source term which we must add to the exponent in the functional 
integral to cancel the linear term after the shift is given by Eq. (\ref{fla13})
with the minus sign. Thus, Feynman diagrams in the external
field (\ref{fla10}) in the light-like gauge ${\cal A}_{*}=0$ are generated
 by the functional integral
\begin{equation}
\!\int\!{\cal D}{\cal A} \exp\Big\{iS({\cal A})+
i\!\int\! d^2x_{\perp} 
U^{ai}(x_{\perp})[{\cal A}^a_i(\infty p_2+x_{\perp})
-{\cal A}^{ai}(-\infty p_2+x_{\perp})]\Big\}
\label{fla14}
\end{equation}
In an arbitrary gauge the source term in the exponent in Eq. (\ref{fla14}) 
can be rewritten in the form
\begin{equation}
2i\int d^2x_{\perp}{\mathop{\rm Tr}} \{U^i(x_{\perp})\int^{\infty}_{-\infty} 
dv[-\infty p_2,vp_2]_{x_{\perp}}
F_{*i}(vp_2+x_{\perp})[vp_2,-\infty p_2]_{x_{\perp}}\}
\label{fla15}
\end{equation}
(Hereafter we use the space-saving notation 
$[up_2,vp_2]_{x_{\perp}}\equiv [up_2+x_{\perp},vp_2+x_{\perp}]$
and similar notation for gauge link ordered along $p_1$). 
Thus, we have found the generating functional for our Feynman diagrams in the 
external field (\ref{fla11}). 

\begin{figure}[htb]
\hspace{3cm}
\mbox{
\epsfxsize=10cm
\epsfysize=7cm
\hspace{0cm}
\epsffile{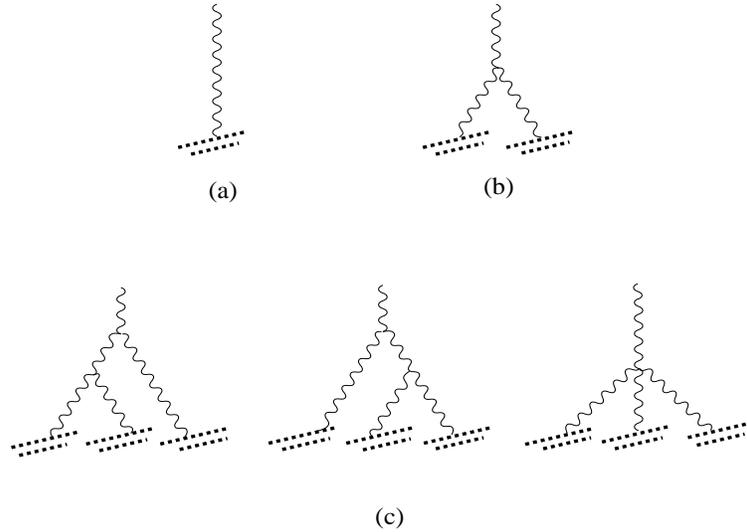}}
\vspace{0cm}
{\caption{Perturbative diagrams for the classical field  (\ref{fla10}) 
 \label{fig6}}}
\end{figure}

It is instructive to see how the source (\ref{fla15}) creates the field
(\ref{fla10}) in perturbation theory. To this end, we
must calculate the field 
\begin{eqnarray}
\bar{\cal A}_{\mu}(x)&=&\!\int\!{\cal D}{\cal A} {\cal A}_{\mu}(x)
\exp\Big\{iS({\cal A})+
2i\!\int d^2x_{\perp}{\mathop{\rm Tr}} \{U^i(x_{\perp})\nonumber\\
&~&\int^{\infty}_{-\infty} 
dv[-\infty p_2,vp_2]_{x_{\perp}}
F_{*i}(vp_2+x_{\perp})[vp_2,-\infty p_2]_{x_{\perp}}\}\Big\}
\label{fla15a}
\end{eqnarray}
by expansion of both 
$S({\cal A})$ and gauge links in the source term (\ref{fla15}) in powers
of $g$ (see Fig. \ref{fig6}). 
In the first order one gets
\begin{equation}
\bar{\cal A}^{(0)}_{\mu}(x)=\int^{\infty}_{-\infty}dv 
\int dz_{\perp}U^{ia}(z_{\perp})\langle{\cal A}_{\mu}(x)
F^a_{*i}(vp_2+z_{\perp})\rangle
\end{equation}
where $\langle{\cal O}\rangle\equiv\int\! {\cal D}{\cal A}e^{iS_0}{\cal O}$.
  Now we must choose a proper gauge for our calculation. We are trying
to create a field (\ref{fla10}) perturbatively and therefore the 
gauge for our perturbative calculation must be compatible with the form
(\ref{fla10}) --- otherwise, we will end up with the gauge rotation of 
the field $B(x)$. 
(For example, in Feynman gauge we
will get the field $\bar{\cal A}_{\mu}$ of the form of the shock wave 
$\bar{\cal A}_i=\bar{\cal A}_{\ast}=0,~
\bar{\cal A}_{\bullet}\sim \delta(x_{\ast})$). It is convenient
to choose the temporal gauge ${\cal A}_0=0$ 
\footnote{The gauge ${\cal A}_{\ast}=0$ 
which we used above is too singular for the perturbative calculation.
In this gauge one must first regulate the external field (\ref{fla10}) 
by, say, replacement 
$U_i\theta(x_{\ast})\rightarrow U_i\theta(x_{\ast})e^{-\epsilon x_{\bullet}}$
and let $\epsilon\rightarrow 0$ only in the final results.}
with the boundary condition $\left.{\cal A}\right|_{t=-\infty}=0$ where
\begin{equation}
{\cal A}_{\mu}(t, \vec{x})=\int^{t}_{-\infty}dt'F_{0\mu}(t',\vec{x})
\label{gau}
\end{equation}
In this gauge we obtain
\begin{eqnarray}
\bar{\cal A}^{(0)}_{\mu}(x)&=&\int\!{dp\over (2\pi)^3}
\left(g_{\mu\nu}-2
{p_{\mu}(p_1+p_2)_{\nu}+(\mu\leftrightarrow\nu)\over 
s(\alpha+\beta+i\epsilon)}+
{4p_{\mu}p_{\nu}\over s^2(\alpha+\beta+i\epsilon)^2}\right)\nonumber\\
&~&
{1\over \alpha\beta s-p_{\perp}^2+i\epsilon}\int dz_{\perp}
e^{i\alpha x_{\bullet}+
i\beta x_{\ast}-i\vec{p}_{\perp}(\vec{x}-\vec{z})_{\perp}}p_{2\nu}
\delta(\alpha{s\over 2})
\partial_iU^{ia}(z_{\perp})
\end{eqnarray}
where $\delta(\alpha{s\over 2})$ comes from the 
$\int dve^{iv\alpha{s\over 2 }}$.  
(Note that the form of the singularity ${1\over(p_0+i\epsilon)}$ which
follops from Eq. (\ref{gau}) differs from conventional Mandelstam-Leibbrandt
prescription $V.p.{1\over p_0}$). Recalling that in terms of 
Sudakov variables 
$dp={s\over 2}d\alpha d\beta dp_{\perp}$ one easily gets that 
$\bar{\cal A}^{(0)}_{\ast}=\bar{\cal A}^{(0)}_{\bullet}=0$ and
\begin{eqnarray}
\bar{\cal A}^{(0)}_{i}(x)&=&\theta(x_{\ast})\int\!{dp\over (2\pi)^2}
{1\over p_{\perp}^2}\int dz_{\perp
}e^{-i\vec{p}_{\perp}(\vec{x}-\vec{z})_{\perp}}
\partial_i
\partial_jU^{ja}(z_{\perp})
\end{eqnarray}
which can be written down formally as
\begin{equation}
-\theta(x_{\ast})
{1\over\partial_{\perp}^2}\partial_i\partial_jU^{j}(x_{\perp})=
U_i(x_{\perp})\theta(x_{\ast})-\theta(x_{\ast}){1\over\partial_{\perp}^2}
(\partial_{\perp}^2g_{ij}+\partial_i\partial_j)U^{j}(x_{\perp})
\end{equation}
(in our notations $\partial_{\perp}^2\equiv-\partial_i\partial^i$).
Now,
since $U_i(x)$ is a pure gauge field (with respect to transverse
coordinates) we have $\partial_i U_j-\partial_j U_i=i[U_i,U_j]$ so
\begin{equation}
\bar{\cal A}^{(0)}_{i}(x)=
U_i(x_{\perp})\theta(x_{\ast})-\theta(x_{\ast})
ig{\partial_j\over\partial_{\perp}^2}
[U_i,U_j])(x_{\perp})
\label{pr7}      
\end{equation}
Thus, we have reproduced the field (\ref{fla10}) up to the correction of of
$g$.   We will demonstrate now that this $O(g)$ correction is canceled
by the next-to-leading term in the expansion of the  exponent of the source
term in eq. (\ref{fla15a}). In the next-to-leading order one gets (see Fig.
\ref{fig6}b): 
\begin{eqnarray}
\lefteqn{\bar{\cal A}^{(1)}_{\mu}(x)=
g\int\!dy\int\!dz_{\perp}dz'_{\perp}U^{ja}(z_{\perp})U^{kb}(z'_{\perp})}
\label{nlo}\\
&\langle{\cal A}_{\mu}(x)
2{\rm Tr}\{\partial_{\alpha}{\cal A}_{\beta}(y)
[{\cal A}_{\alpha}(y),{\cal A}_{\beta}(y)]\}
\int\! dv F^a_{*j}(vp_2+z_{\perp})
\int\! dv'F^b_{*k}(vp_2+z'_{\perp})\rangle\nonumber
\end{eqnarray}
It is easy to see that 
$\bar{\cal A}^{(1)}_{\ast}=\bar{\cal A}^{(1)}_{\bullet}=0$
and
\begin{eqnarray}
\lefteqn{\bar{\cal A}^{(1)}_{i}(x)=}\\
&g\int\!dy\int{dp\over (2\pi)^4i}e^{-ip(x-y)}{1\over p^2}
\left(\partial^k[{\cal A}^{(0)}_{i}(y),{\cal A}^{(0)}_{k}(y)]+
[{\cal A}^{(0)k}(y),\partial_i{\cal A}^{(0)}_{k}(y)-(
i\leftrightarrow k)]\right) \nonumber 
\end{eqnarray}
Since ${\cal A}^{(0)}_{k}$ is given by Eq. (\ref{pr7}) this reduces to
\begin{eqnarray}
\bar{\cal A}^{(1)}_{i}(x)=-g\theta(x_{\ast})\!\int\!dy_{\perp}{dp_{\perp}\over
(2\pi)^2} {e^{-ip_{\perp}(x-y)_{\perp}}\over p_{\perp}^2}
i\partial^k([U_{i}(y),U_{k}(y)])+O(g^2)
\end{eqnarray}
which cancels the second term in Eq. (\ref{pr7}). Thus, we obtain
\begin{equation}
\bar{\cal A}_{i}(x)=
U_i(x_{\perp})\theta(x_{\ast})+O(g^2)
\label{pr8}      
\end{equation}
Similarly, one can check that the contributions $\sim g^2$ coming the diagrams 
in Fig. \ref{fig6}c cancel the $g^2$ term in the Eq. (\ref{pr8}) and so
on leading finally to the expression $U_i(x_{\perp})\theta(x_{\ast})$ without 
any corrections. 

\begin{figure}[htb]
\hspace{5cm}
\mbox{
\epsfxsize=3.5cm
\epsfysize=6cm
\hspace{0cm}
\epsffile{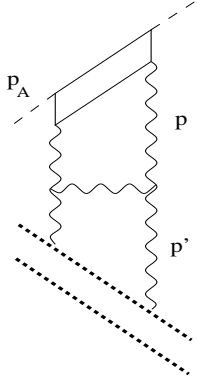}}
\vspace{-0.5cm}
{\caption{A typical loop diagram in the external field 
created by the Wilson-line 
source (\ref{fla15}) \label{fig7}}}
\end{figure}

We have found the generating functional 
for the diagrams in the external field (\ref{fla10}) which give the coefficient
functions in front of our Wilson-line operators $U_i$.
Note that formally we obtained the source term with the gauge link 
ordered along the light-like line which is a potentially dangerous situation. 
Indeed, it it is easy to see that already the first loop diagram shown
in Fig. \ref{fig7} is divergent. 
The reason is that the longitudinal integrals
over  $\alpha_p$ are unrestricted from below (if the Wilson 
line is light-like).
However,  this is not what we want for the
coefficient functions because they should include only the integration over
the  region $\alpha_p>\sigma$ (the region $\alpha_p<\sigma$ 
belongs to matrix elements, see the discussion in Sect. 3). Therefore,
we must impose somehow this condition $\alpha_p>\sigma$ in our Feynman diagrams
created by the source (\ref{fla15}). Fortunately, we already faced similar
problem --- how to impose a condition $\alpha_p<\sigma$  on the matrix
elements of operators $U$ (see Fig. \ref{fig4}) 
and we have solved that problem by changing the slope of the supporting
 line. We demonstrated that in order to cut the integration over 
 large $\alpha>\sigma$ 
from matrix elements of Wilson-line operators $U_i$  we need 
to change the slope of these Wilson-line operators
to  $n=\sigma p_1+\tilde{\sigma} p_2$. Similarly,  
if we want to cut the integration over
small $\alpha_p<\sigma$ from the
coefficient functions we need to order the gauge 
factors in  Eq.(\ref{fla15}) along (the same) vector 
$n=\sigma p_1+\tilde{\sigma} p_2$
\footnote{
Note that the 
diagram in Fig. \ref{fig7} is the diagram in Fig. \ref{fig4}b turned
upside down.  In the Fig. \ref{fig4}b diagram we have a restriction
$\alpha<\sigma$. It is easy to see that 
this  also means a restriction $\beta>\tilde{\sigma}$ 
if one chooses to write down the rapidity 
integrals in terms of $\beta$'s rather than $\alpha$'s. Turning
the diagram upside down amounts to interchange of $p_A$ and $p_B$ which
leads to ({\bf i}) replacement of the slope of Wilson line by 
$\tilde{\sigma}p_1+\sigma p_2$ and ({\bf ii})
replacement $\alpha\leftrightarrow \beta$ in the integrals. 
Thus, the restriction
$\beta>\tilde{\sigma}$ imposed by the line collinear to 
$\sigma p_1+\tilde{\sigma} p_2$ in diagram in Fig. \ref{fig4}b means
the restriction $\alpha>\tilde{\sigma}$ by the line 
$\parallel~\tilde{\sigma}p_1+\sigma p_2$ in the Fig. \ref{fig7} diagram. 
After renaming $\sigma$ by $\bar{\sigma}$ we obtain 
 the desired result.}.
 
Therefore, the final form of
the generating functional for the Feynman diagrams (with  $\alpha>\sigma$
cutoff) in the external field  (\ref{fla11}) is 
\begin{eqnarray} \int\!{\cal
D}{\cal A} {\cal D}\Psi {\rm exp}\left\{iS({\cal A},\Psi)+ i\int d^2x_{\perp}
U^{ai}(x_{\perp})V^a_i(x_{\perp})\right\} \label{fla16} 
\end{eqnarray} 
where
\begin{eqnarray} \lefteqn{V_i(x_{\perp})=}\label{fla17}\\
&\int^{\infty}_{-\infty} dv[-\infty n,vn]_{x_{\perp}}n^{\mu}F_{\mu
i}(vn+x_{\perp}) [vn,-\infty n]_{x_{\perp}}
\nonumber
\end{eqnarray}
and $V^a_i\equiv {\rm Tr}(\lambda^aV_i)$ as usual. For completeness, 
we have added 
integration over quark fields so $S({\cal A},\Psi)$ is the full QCD action.
 
 Now we can assemble the different parts of the factorization 
formula (\ref{fla4}). We have written down the generating functional integral 
for the diagrams with $\alpha>\sigma$ in the external fields with 
$\alpha<\sigma$ and what remains now is to write down the integral over 
these ``external'' fields. 
Since this
integral is completely independent of (\ref{fla16}) we will use a different
notation ${\cal B}$ and $\chi$ for the $\alpha<\sigma$ fields. We have: 
\begin{eqnarray}
\lefteqn{\int\!\! {\cal D}A{\cal D}\bar{\Psi}{\cal D}\Psi e^{iS(A,\Psi)} 
j(p_{A})j(p'_{A})j(p_{B})j(p'_{B})=}
\label{fla18}\\ 
&\int\!\! {\cal D}{\cal A}{\cal D}\bar{\psi}{\cal D}
\psi e^{iS({\cal A},\psi)} j(p_{A})j(p'_{A})
\int\!\! {\cal D}{\cal B}{\cal D}\bar{\chi}{\cal D}\chi \nonumber\\
&
j(p_{B})j(p'_{B})
e^{iS({\cal B},\chi)} \exp\Big\{i\!\int\! d^2x_{\perp} 
U^{ai}(x_{\perp})V^{a}_i(x_{\perp})\Big\}\nonumber
\end{eqnarray}
The operator $U_i$ in an arbitrary gauge is
given by the same formula (\ref{fla17}) as operator $V_i$
with the only difference that the gauge links and $F_{{\bullet} i}$ 
are constructed from the fields 
${\cal B}_{\mu}$. This is our factorization formula (\ref{fla4}) 
in the functional integral representation.

The functional integrals over ${\cal A}$ fields give logarithms of the
type $g^2\ln{1/\sigma}$ while the integrals over slow ${\cal B}$ fields give
powers of $g^2\ln (\sigma s/m^2)$. With logarithmic accuracy, they add up to
$g^2\ln s/m^2$. However, there will be
additional terms $\sim g^2$ due to mismatch coming from the region 
of integration near the dividing point $\alpha\sim\sigma$ where the
details of the cutoff in the matrix elements of the operators $U$ and $V$ 
become important. Therefore, one should expect the corrections of order of 
$g^2$ to the effective action $\int dx_{\perp} U^iV_i$. Still,
the fact that the fast quark moves along the straight line has nothing
to do with perturbation theory (cf. ref. \cite{dosch}); therefore it is 
natural to expect the
non-perturbative generalization of the factorization formula (\ref{fla18}) 
constructed from the same Wilson-line operators $U_i$ and $V_i$
(probably with some kind of non-local interactions between them).

\section{Effective action for high-energy scattering}
The factorization formula gives us a starting point for a new approach
to the analysis of the high-energy effective action. 
Consider another rapidity $\eta'_0$ in the region between $\eta_0$ and 
$\eta_B=\ln m^2/s$. If we use the factorization formula 
(\ref{fla18}) once more, this time dividing between the rapidities 
greater and smaller than $\eta'_0$, we get the expression 
for the amplitude (\ref{fla5}) in the form
\footnote
{For brevity, we do not display the quark fields.}:
\begin{eqnarray}
iA(s,t)&=&\int\!\! {\cal D}Ae^{iS(A)} 
j(p_{A})j(p'_{A})j(p_{B})j(p'_{B})
\label{fla19}\\ 
&=&\int\!\! {\cal D}{\cal A} e^{iS({\cal A})} j(p_{A})j(p'_{A})
\int\!\! {\cal D}{\cal B}
e^{iS({\cal B})}j(p_{B})j(p'_{B}) \nonumber\\
&~&\int\! {\cal D}{\cal C}e^{iS({\cal C})}
e^{i\!\int\! d^2x_{\perp} 
V^{ai}(x_{\perp})U^a_i(x_{\perp})+i\!\int\! d^2x_{\perp} 
W^{ai}(x_{\perp})Y^a_i(x_{\perp})}
\nonumber
\end{eqnarray}
In this formula the operators $V_i$ (made from ${\cal A}$ fields) 
are given by Eq. (\ref{fla17}), the operators $U_i$ are also given by 
Eq. (\ref{fla17}) but constructed from the 
${\cal C}$ fields instead, and the operators $W_i$ (made from 
${\cal C}$ fields) and $Y_i$ 
(made from ${\cal B}$ fields) are aligned along the 
direction $n'=\sigma'p_1+\tilde{\sigma}'p_2$ 
corresponding to the rapidity $\eta'$ (as usual, 
$\ln \sigma'/\tilde{\sigma}'=\eta'$ 
where
$\tilde{\sigma}'=m^2/s\sigma'$):
\begin{eqnarray}
&U_i({\cal C})_{x_{\perp}}=\int^{\infty}_{-\infty} 
dv[-\infty n,vn]_{x_{\perp}}
n^{\mu}F_{\mu i}(vn+x_{\perp})
[vn,-\infty n]_{x_{\perp}}\nonumber\\
&W_i({\cal C})_{x_{\perp}}=\int^{\infty}_{-\infty} 
dv[-\infty n',vn']_{x_{\perp}}
n^{'\mu}F_{\mu i}(vn'+x_{\perp})
[vn',-\infty n']_{x_{\perp}}\nonumber\\
&Y_i({\cal B})_{x_{\perp}}=\int^{\infty}_{-\infty} 
dv[-\infty n',vn']_{x_{\perp}}
n^{'\mu}F_{\mu i}(vn'+x_{\perp})
[vn',-\infty n']_{x_{\perp}}
\nonumber
\end{eqnarray}
Thus, we have factorized the functional integral 
over ``old'' ${\cal B}$ fields 
into the product of two integrals over ${\cal C}$ and ``new" ${\cal B}$
fields.

Now, let us integrate over the ${\cal C}$ fields  
and write down the result in terms of an effective action. 
Formally, one obtains:
\begin{equation}
iA(s,t)=
\int\!\! {\cal D}{\cal A} e^{iS({\cal A})} j(p_{A})j(p'_{A})
\int\!\! {\cal D}{\cal B}
e^{iS({\cal B})}j(p_{B})j(p'_{B})
e^{iS_{\rm eff}(V_i,Y_i;{\sigma\over\sigma'})} 
\label{fla21}
\end{equation}
where $S_{\rm eff}$ for the rapidity interval between $\eta$ and 
$\eta'$ is defined as 
\begin{eqnarray}
&e^{iS_{\rm eff}(V_i,Y_i;{\sigma\over\sigma'})}=
\int\! {\cal D}{\cal C}e^{iS({\cal C})}
e^{i\!\int\! d^2x_{\perp} 
V^{ai}(x_{\perp})U^a_i(x_{\perp})+i\!\int\! d^2x_{\perp} 
W^{ai}(x_{\perp})Y^a_i(x_{\perp})}\label{fla22}
\end{eqnarray}
This formula gives a rigorous definition for the effective action for a 
given interval in rapidity 
(cf. ref. \cite{lobzor}).

\begin{figure}[htb]
\hspace{1cm}
\mbox{
\epsfxsize=12cm
\epsfysize=4cm
\hspace{0cm}
\epsffile{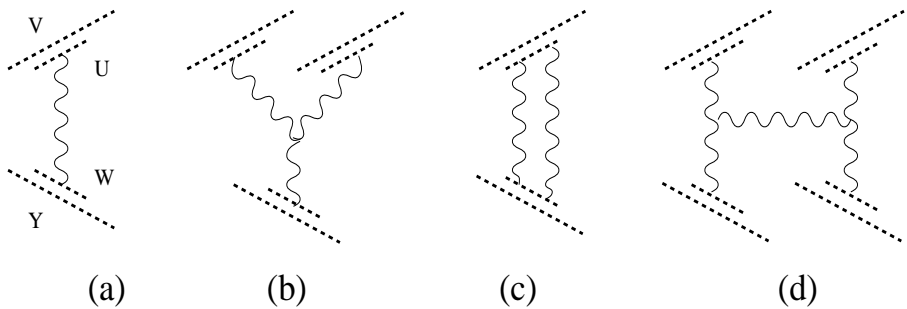}}
\vspace{0cm}
{\caption{Lowest order terms in the perturbative expansion 
of the effective action. \label{fig8}}}
\end{figure}

\begin{figure}[htb]
\hspace{0cm}
\mbox{
\epsfxsize=16cm
\epsfysize=5cm
\hspace{0cm}
\epsffile{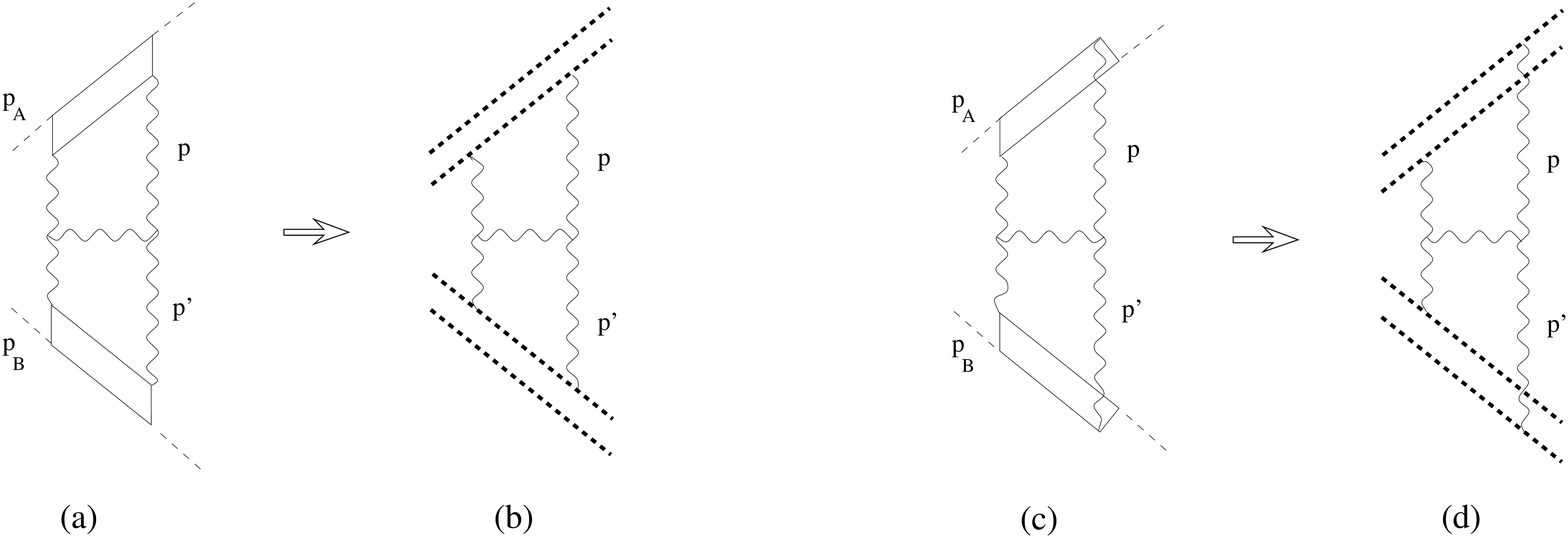}}
\vspace{0.5cm}
{\caption{Counting of loops for Feynman diagrams (a),(c) and the corresponding
Wilson-line operators (b),(d)\label{fig4a}}}
\end{figure}

Next step would be to perform explicitly the integrations over the 
longitudinal momenta in the r.h.s. 
of Eq. (\ref{fla22}) and obtain the answer
for the integration over
our rapidity region (from $\eta$ to $\eta'$) in terms of two-dimensional 
theory in the transverse coordinate space which hopefully would give us 
the unitarization of the BFKL pomeron. At present, it is not 
known how to do this. One can obtain, however, a first few terms in the 
expansion of effective action in powers of $V_i$ and $Y_i$. The easiest way 
to do this is to expand gauge factors $U_i$ and $W_i$ in r.h.s. of Eq. 
(\ref{fla22}) in powers of ${\cal C}$ fields and calculate the relevant 
perturbative diagrams (see Fig. \ref{fig8}).
The first few terms in the effective action at the one-log level 
\footnote{
This "one-log" level corresponds to one-loop level for usual Feynman
diagrams.
Superficially, the diagram in Fig. \ref{fig8}d looks like tree diagram in
comparison to diagram in Fig. \ref{fig8}c which has one loop. 
However, both of the diagrams in Fig. \ref{fig8}c and d  contain 
integration over longitudinal momenta
(and thus the factor $\ln{\sigma\over\sigma'}$) so in the longituduinal
space the diagram in Fig. \ref{fig8}d is a loop diagram too. 
It happens because for diagrams with 
Wilson-line operators the 
counting of number of loops literally corresponds to the counting of the 
number of loop integrals only for the transverse momenta. For
the longitudinal variables, the diagrams which look like 
trees may contain logarithmical loop integrations. This property is 
illustrated in Fig. \ref{fig4a}: the Wilson-line diagram shown in 
Fig. \ref{fig4a}b has two loops and the diagram shown in Fig. \ref{fig4a}d  
is a tree but both of them originated from  Feynman diagrams 
shown in Fig. \ref{fig4a}a and c with equal number of loops. 
To avoid confusion, we will use the termin ``one-log 
level" instead of "one-loop level".}
have the form \cite{verlinde},\cite{many}:
\begin{eqnarray}
\lefteqn{S_{\rm eff} = \int d^2x 
V^{ai}(x)Y^a_i(x)-}\label{fla23}\\
&{g^2\over 64\pi^3}\ln{\sigma\over\sigma'}
\Big(N_c\int d^2x d^2y V^a_{i,i}(x)\ln^2(x-y)^2Y^a_{j,j}(y)+
{f_{abc}f_{mnc}\over 4\pi^2}\!\int\! d^2x d^2y d^2x' d^2y'd^2z\nonumber\\
&
V^a_{i,i}(x)V^m_{j,j}(y)Y^b_{k,k}(x')Y^n_{l,l}(y')
\ln{(x-z)^2\over(x-x')^2}\ln{(y-z)^2\over(y-y')^2}
\left({\partial\over\partial z_i}\right)^2\ln{(x'-z)^2\over(x-x')^2}
\ln{(y'-z)^2\over(y-y')^2}\Big) +...
\nonumber
\end{eqnarray}
where we we use the notation 
$V^a_{i,j}(x)\equiv {\partial\over \partial x_j}V^a_i(x)$ etc. 
The first term (see Fig. \ref{fig8}a) looks like the corresponding term in the 
factorization 
formula (\ref{fla18}) -- only the directions of the supporting lines are 
now strongly different 
\footnote{
Strictly speaking, the contribution coming from the diagram shown in 
Fig. \ref{fig8}a
has the form 
$\int d^2x V^{ai}(x){\partial_i\partial_j\over\partial^2}Y^{aj}(x)$
which differs from the first term in r.h.s. of eq. (\ref{fla23}) by
$\int d^2x V^{ai}(x){1\over\partial^2}
(\partial^2g_{ij}-\partial_i\partial_j)Y^{aj}(x)$. However, 
 it may be demonstrated that this discrepancy 
(which is actually
$\sim O(g)$ for a a pure gauge field $Y_i$) is canceled by the 
contribution from the diagram with three-gluon vertex shown in Fig. \ref{fig8}b 
just
as in the case of perturbative calculation of ${\cal A}_i$ discussed in
Sect.3.} .  
The second
term shown in Fig. \ref{fig8}c is the first-order expression for the 
reggeization of the gluon\cite{bfkl}
and the third term (see Fig. \ref{fig8}d) is the two-reggeon  
Lipatov's Hamiltonian\cite{l} responsible for BFKL logarithms. 

\begin{figure}[htb]
\hspace{3cm}
\mbox{
\epsfxsize=8cm
\epsfysize=4cm
\hspace{0cm}
\epsffile{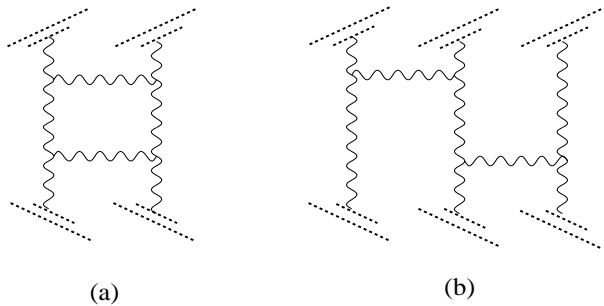}}
\vspace{0cm}
{\caption{Typical perturbative diagrams in the next
$\left(\ln{\sigma\over\sigma'}\right)^2$ order.\label{fig9}}}
\end{figure}

Let us discuss subsequent terms in the perturbative expansion (\ref{fla23}).
There can be two types of the logarithmical contributions. First is
the "true" loop contribution coming from the diagrams of the Fig.\ref{fig9}a 
type. This diagram is an iteration of the Lipatov's Hamiltonian. However, 
in the same $(\ln{\sigma\over\sigma'})^2$ order there is another 
contribution
coming from the diagram shown in Fig. \ref{fig9}b. 
To treat them separately, we can consider the case
when $g\ll 1$ but the sources are strong ($\sim {1\over g}$) 
so $gY_i\sim gU_i\sim 1$. In this case, the diagram in Fig.\ref{fig9}a
has the order 
$g^4Y_i^2V_i^2\left(\ln{\sigma\over\sigma'}\right)^2
\sim \left(\ln{\sigma\over\sigma'}\right)^2$ while the "tree" 
Fig.\ref{fig8}b diagram is 
$\sim g^4Y_i^3V_i^3\left(\ln{\sigma\over\sigma'}\right)^2
\sim {1\over g^2}\left(\ln{\sigma\over\sigma'}\right)^2$. So,
in this approximation the tree diagrams are the most important   
and should be summed up in the first place. As usual, the best way to sum 
the tree diagrams is given by the semiclassical
method which will be discussed in next Section.
 
\begin{figure}[htb]
\hspace{3cm}
\mbox{
\epsfxsize=10cm
\epsfysize=10cm
\epsffile{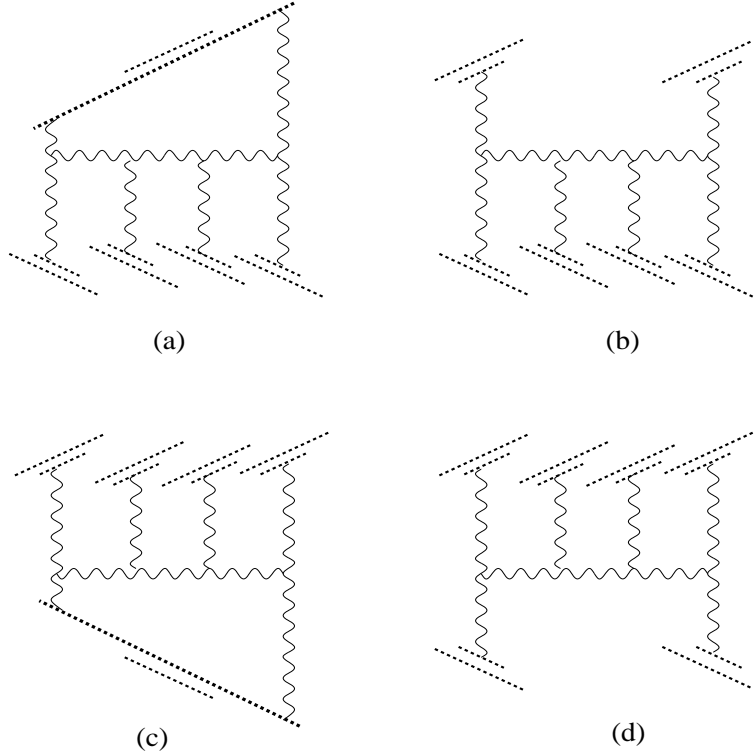}}
\vspace{0.5cm}
{\caption{Perturbative diagrams for the effective action 
in the case of one weak source and one strong one.\label{fig10}}}
\end{figure}

 However, if we would like to get the result on the one-log level 
 it can be obtained using the evolution equations 
 for the Wilson-line operators \cite{ing}. Note that at this level 
 we have only the diagrams of the Fig.\ref{fig10} type. 
 These diagrams describe the 
 situation when one of the sources is weak and another is still strong
 (see also refs. \cite{larry2}, \cite{kovner}).
 For example, if the source $V_i$ is weak 
 (and hence $gV_i$ is a valid small parameter) but the source $Y_i$ is 
 not weak (so that $gV_i\sim 1$ is {\it not} a small parameter) one must 
 take
 into account the diagrams shown in Fig. \ref{fig10}a and b.
The multiple rescatterings in Fig. \ref{fig10}a,b  describe the motion of the 
gluon emitted by the 
weak source $V_i$ in the strong external field $A_i=Y_i\theta(x_{\ast})$
created by the source $Y_i$. These diagrams were calculated in ref. \cite{ing}.
For example, the result of the calculation of the diagram in Fig. 
\ref{fig10}a presented in a form of the evolution of the 
Wilson-line operators $U_i$ reads
\footnote{Here 
$Y_x\equiv Y(x_{\perp})=[\infty n'+x_{\perp}, -\infty n'+x_{\perp}]$ so that
$Y_i(x_{\perp})=Y^{\dagger}_xi{\partial\over\partial_i}Y_x$. (Note that
 we have the gauge factors in the gluon (adjoint) representation here).
 }
\begin{eqnarray}
\lefteqn{U^a_i(x_{\perp})\rightarrow Y^a_i(x_{\perp})-}\nonumber\\
&
{g^2\over 8\pi^3}\ln{\sigma\over\sigma'}\int dy_{\perp}{1\over(x-y)_{\perp}^2}
\Big(f^{abc}\big(Y^{\dagger}_x\partial_iY_y\big)^{bc}+
N_cU^a_i(x_{\perp})\Big)+...
\label{dob1}
\end{eqnarray}
where dots stand for the terms with extra 
$g^2\ln{\sigma\over\sigma'}$ factors. This evolution equation 
means that if we integrate over the rapidities $\eta_0>\eta>\eta'_0$  
in the matrix elements of the 
operator $U_i$ we will get the expression (\ref{dob1}) constructed 
from the operators $Y_i$ with rapidities up to $\eta'_0$ times 
 factors proportional to  
 $g^2(\eta_0-\eta'_0)\equiv g^2\ln{\sigma\over\sigma'}$.
Therefore, the corresponding contribution to the effective action 
at the one-log level takes the form
\begin{eqnarray}
\lefteqn{\int dx_{\perp}V^a_i(x_{\perp}) U^{ai}(x_{\perp})\rightarrow 
\int dx_{\perp} V^a_i(x_{\perp})Y^{ai}(x_{\perp})-}\label{dob2}\\
&
{g^2\over 8\pi^3}\ln{\sigma\over\sigma'}\int\! dx_{\perp} dy_{\perp}
{1\over(x-y)_{\perp}^2}
\Big(i\big(V^{i}(x_{\perp})Y^{\dagger}_x\partial_iY_y\big)^{aa}-
N_cV^{ai}(x_{\perp})U^a_i(x_{\perp})\Big)
\nonumber
\end{eqnarray}
where the first term is the lowest-order 
effective action ($\equiv$ the first term in eq. (\ref{fla23})) and 
the second term contains new information. 
To check this second term, we may expand it in
powers of the source $Y_i$ and it is easy to see that 
the first nontrivial term in this expansion coincides with the 
gluon-reggeization term in eq. (\ref{fla23}). 

Apart from the (\ref{dob2}) term, 
there is another contribution to the one-loop evolution equations 
coming from the diagrams in Fig. (\ref{fig10}b) \cite{ing}:
\begin{eqnarray}
\lefteqn{U_i^a(x_{\perp})U_j^b(y_{\perp})\rightarrow}\label{dob3}\\ 
&-{g^2\over 4\pi^3}\ln{\sigma\over\sigma'}
\Big(\nabla^x_i\Big[\int dz_{\perp}
{(x-z)_{\perp}\cdot(y-z)_{\perp}\over(x-z)^2_{\perp}(y-z)^2_{\perp}}
(Y^{\dagger}_xY_y+1-Y^{\dagger}_xY_z- Y^{\dagger}_zY_y)\Big]
\stackrel{\leftarrow}{\nabla}^y_j\Big)^{ab}
\nonumber
\end{eqnarray}
where
\begin{eqnarray}
\nabla^x_i{\cal O}(x_{\perp})&\equiv& 
{\partial\over\partial x^i}{\cal O}(x_{\perp})-
iU_i(x_{\perp}){\cal O}(x_{\perp})\nonumber\\
{\cal O}(y_{\perp})\stackrel{\leftarrow}{\nabla}^y_i&\equiv& 
-{\partial\over\partial y^i}{\cal O}(y_{\perp})-
i{\cal O}(y_{\perp})U_i(y_{\perp})
\label{dob4}
\end{eqnarray}
are the ``covariant derivatives" (in the adjoint representation). 
The corresponding term in effective action has the form
\begin{eqnarray}
&&{ig^2\over 8\pi^3}\ln{\sigma\over\sigma'}
\int dx_{\perp}dy_{\perp} \left(\nabla^x_iV^a_i\right)(x_{\perp})
\int dz_{\perp}
{(x-z)_{\perp}\cdot(y-z)_{\perp}\over(x-z)^2_{\perp}(y-z)^2_{\perp}}
\nonumber\\
&&
\big(Y^{\dagger}_xY_y+1-Y^{\dagger}_xY_z- Y^{\dagger}_zY_y\big)^{ab}
\left(\nabla^y_jV^b_j\right)(y_{\perp})
\label{dob5}
\end{eqnarray}
The final form of the one-log effective action for this case 
is the sum of the expressions 
(\ref{dob2}) and (\ref{dob5}):
\begin{eqnarray}
\lefteqn{S^{(I)}_{\rm eff}(V_i,Y_j) = \int d^2x 
V^{ai}(x)Y^a_i(x)-}\label{dob6}\\
&
{g^2\over 8\pi^3}\ln{\sigma\over\sigma'}\int\! dx_{\perp} dy_{\perp}
{1\over(x-y)_{\perp}^2}
\Big(i\big(V^{i}(x_{\perp})Y^{\dagger}_x\partial_iY_y\big)^{aa}-
N_cV^{ai}(x_{\perp})U^a_i(x_{\perp})\Big)
\nonumber\\
&+{ig^2\over 8\pi^3}\ln{\sigma\over\sigma'}\int dx_{\perp}dy_{\perp}
\nabla^x_iV^{ai}(x_{\perp})\int dz_{\perp}
{(x-z)_{\perp}\cdot(y-z)_{\perp}\over(x-z)^2_{\perp}(y-z)^2_{\perp}}
\nonumber\\
&\big(Y^{\dagger}_xY_y+1-Y^{\dagger}_xY_z- Y^{\dagger}_zY_y\big)^{ab}
\nabla^y_jV^{bj}(y_{\perp}) 
\nonumber
\end{eqnarray}
where $V_i$ is a weak source and $Y_i$ is
a strong one. It is clear that if 
the source $V_i$ is strong and $Y_i$ is weak as shown in Fig. \ref{fig9}c,d 
diagrams the effective action $S^{(II)}_{\rm eff}(V_i,Y_j)$
will have the similar form with the replacement $V\leftrightarrow Y$.

As we mentioned above, the diagrams in Fig.\ref{fig9} and Fig. \ref{fig10}
complete the list of diagrams which contribute to the effective action at 
the one-log level (even if both sources are strong). It means that the one-log
answer in general case can be guessed by comparison of the answers for 
$S^{(I)}_{\rm eff}(V_i,Y_j)$ and
$S^{(II)}_{\rm eff}(V_i,Y_j)$ (the simple sum is not enough since 
some of the contributions will be double-counted). Instead of doing that,
we will obtain the one-log result for two strong sources 
using the semiclassical method and check that it agrees with (\ref{dob6}).

\section{Effective action and collision of two shock waves} 

The functional integral (\ref{fla22})
which defines the effective action is the usual QCD functional integral 
with two sources corresponding to the two colliding shock waves. 
Instead of calculation of perturbative diagrams 
(as it was done in previous section) one can use the
semiclassical approach. This approach
is relevant when the coupling constant is relatively small but the 
characteristic fields are large (in other words, when $g^2\ll 1$ 
but $gV_i\sim gY_i \sim 1$). In this case one can 
calculate the functional integral (\ref{fla22}) by expansion around the new
stationary point corresponding to the classical
wave created by the collision of the shock waves.

With leading log accuracy, we can replace the vector $n$ by $p_1$ and the
vector $n'$ by $p_2$. Then the functional integral (\ref{fla22}) 
takes the  form:
\begin{eqnarray}
&e^{iS_{\rm eff}(V_i,Y_i;{\sigma\over\sigma'})}=
\int\! {\cal D}Ae^{iS_{QCD}(A)}
e^{i\!\int\! d^2x_{\perp} 
V^{ai}(x_{\perp})U^{a}_i(x_{\perp})
+i\!\int\! d^2x_{\perp}W^{ai}
Y^a_i(x_{\perp})}\label{fla24}
\end{eqnarray}
where now
\begin{equation}
U^{a}_i(x_{\perp})=\int^{\infty}_{-\infty} dv
\hat{F}_{\bullet i}(vp_1+x_{\perp}),~~~~~~~
W^{a}_i=\int^{\infty}_{-\infty} dv
\tilde{F}_{\ast i}(vp_2+x_{\perp})
\label{fla25}
\end{equation}
Hereafter we use the notations 
\begin{eqnarray}
\hat{\cal O}(x)&=&
[-\infty p_1+x,x]{\cal O}(x)[x,-\infty p_1+x]\nonumber\\
\tilde{\cal O}(x)&=&
[-\infty p_2+x,x]{\cal O}(x)[x,-\infty p_2+x]
\label{fla26}
\end{eqnarray}
Note that we changed the name for the gluon fields in the integrand 
from ${\cal C}$ back to $A$.

As usual, the classical equation for the saddle point $\bar{A}$ in the 
functional integral (\ref{fla24}) is
\begin{equation}
\left.{\delta \over \delta A}\left(S_{QCD}+\!\int\! d^2x_{\perp} 
V^{ai}(x_{\perp})U^{a}_i(x_{\perp})
+\!\int\! d^2x_{\perp}W^{ai}Y^a_i(x_{\perp}\right)\right|_{A=\bar{A}}=0
\label{fla27}
\end{equation}
To write down them explicitly we need the first variational 
derivatives of the source terms with respect to gauge field. 
We have:
\begin{eqnarray}
&\delta U_i=\delta \hat{A}_i(\infty p_1+x_{\perp})-
\delta {A}_i
(-\infty p_1+x_{\perp})-
\int^{\infty}_{-\infty}du
\hat{\nabla}_i\delta\hat{A}_i(u p_1+x_{\perp})\nonumber\\
&\delta W_i=\delta \tilde{A}_i(\infty p_2+x_{\perp})-
\delta {A}_i
(-\infty p_2+x_{\perp})-
\int^{\infty}_{-\infty}du
\tilde{\nabla}_i\delta\tilde{A}_i(u p_2+x_{\perp})
\label{fla28}
\end{eqnarray}
where 
\begin{eqnarray}
\hat{\nabla}_i{\cal O}(x)&\equiv&\partial_i{\cal O}(x)
-i[U_i(x_{\perp})+A_i(-\infty p_1+x_{\perp}),{\cal O}(x)]
\nonumber\\
\tilde{\nabla}_i{\cal O}(x)&\equiv&\partial_i{\cal O}(x)
-i[W_i(x_{\perp})+A_i(-\infty p_2+x_{\perp}),{\cal O}(x)]
\label{fla29}
\end{eqnarray}
Therefore the explicit form of the classical equations (\ref{fla27}) 
for the wave 
created by the collision is:
\begin{eqnarray}
\lefteqn{D^{\mu}F_{\mu i}=0}\label{fla48}\\
&D^{\mu}F_{\ast\mu}=
\delta({2\over s}x_{\bullet})[{2\over s}x_{\ast}p_1, 
-\infty p_1]_{x_{\perp}}\hat{\nabla}_i V^i(x_{\perp})
[-\infty p_1,{2\over s}x_{\ast}p_1]_{x_{\perp}}\nonumber\\
&D^{\mu}F_{\bullet\mu}=
\delta({2\over s}x_{\ast})[{2\over s}x_{\bullet}p_2, 
-\infty p_2]_{x_{\perp}}\tilde{\nabla}_i Y^i(x_{\perp})
[-\infty p_2,{2\over s}x_{\bullet}p_2]_{x_{\perp}}
\nonumber
\end{eqnarray}

Also, as explained in Sect. 3, since our fields do not decrease at
infinity there may be extra surface linear terms (cf. Eq. (\ref{fla12})) 
coming from the contributions proportional to $\delta A(\pm\infty)$ in 
the r.h.s. of eq. (\ref{fla28}). 
The requirement
of absence of such terms gives four additional equations 
\begin{eqnarray}
\left.F_{\bullet i}\right|_{x_{\bullet}=\infty}&=&
\delta(2x_{\ast}/s)Y_i(x_{\perp}),~~~~~~~
\left.F_{\ast i}\right|_{x_{\ast}=-\infty}= \delta(2x_{\bullet}/s)
V_i(x_{\perp}),\label{fla49}\\ 
\left.F_{\bullet i}\right|_{x_{\bullet}=\infty}&=&\delta(2x_{\ast}/s)
[\infty p_2,- \infty p_2]_{x_{\perp}} Y_i(x_{\perp}) [-\infty
p_2, \infty p_2]_{x_{\perp}}\nonumber\\ 
\left.F_{\ast i}\right|_{x_{\ast}=\infty}&=&\delta(2x_{\bullet}/s)
[\infty p_1, -\infty p_1]_{x_{\perp}} V_i(x_{\perp})
[-\infty p_1, \infty p_1]_{x_{\perp}}
\nonumber
\end{eqnarray}
The two sets of equations (\ref{fla48}) and (\ref{fla49})
define the classical field created by the collision of two shock waves
\footnote{These equations are essentially equivalent to the classical
 equations 
describing the collision of two heavy nuclei in ref. \cite{kovner}.}.

Unfortunately, it is not clear how to solve these equations. One can 
start with the trial field
which is a simple superposition of the two shock waves (\ref{fla10}) 
\begin{equation}
A^{(0)}_{\ast}=A^{(0)}_{\bullet}=0,~~~~
A^{(0)}_i=\Theta(x_{\bullet})V_i+\Theta(x_{\ast})Y_i
\label{fla50}
\end{equation}
and improve
it by taking into account the interaction between the shock waves order by
order\cite{tok}. The parameter of this expansion is the 
commutator $g^2[Y_i,V_k]$. Moreover,  it can be demonstrated that 
each extra commutator brings a factor
$\ln{\sigma\over\sigma'}$ and therefore this approach is a sort of
leading logarithmic approximation.  In the lowest nontrivial order one gets:
\begin{eqnarray} 
A^{(1)}_i&=&-{g\over 4\pi^2}\int
dz_{\perp}([Y_i(z_{\perp}),V_k(z_{\perp})]-i\leftrightarrow k)
{(x-z)^k\over(x-z)_{\perp}^2}\ln\left(1-{(x-z)_{\perp}^2\over
x_{\parallel}^2+i\epsilon}\right) \nonumber\\ 
A^{(1)}_{\bullet}&=&{gs\over 16\pi^2}\int
dz_{\perp} {1\over
x_{\ast}+i\epsilon}\ln(-x_{\parallel}^2+(x-z)_{\perp}^2+i\epsilon)
[Y_k(z_{\perp}),V^k(z_{\perp})] \nonumber\\ 
A^{(1)}_{\ast}&=&-{gs\over 16\pi^2}\int dz_{\perp}
{1\over x_{\bullet}+i\epsilon}\ln(-x_{\parallel}^2+(x-z)_{\perp}^2+i\epsilon)
[Y_k(z_{\perp}),V^k(z_{\perp})] \label{fla51}
\end{eqnarray}
where $x_{\parallel}^2\equiv{4\over s}x_{\ast}x_{\bullet}$ is a 
longitudinal part of $x^2$. These fields are
obtained in the background-Feynman gauge. The corresponding expressions for
field strength have the form:
\begin{eqnarray}
F^{(1)}_{\bullet\ast}&=&{gs\over 4\pi^2}
\int dz_{\perp}{1\over -x_{\parallel}^2+(x-z)_{\perp}^2+i\epsilon}[Y_k,V^k]
\label{fla52}\\
F^{(1)}_{ik}&=&{g\over 2\pi^2}
\int dz_{\perp}{1\over -x_{\parallel}^2+(x-z)_{\perp}^2+i\epsilon}
([Y_i,V_k]-[Y_k,V_i])\nonumber\\
F^{(1)}_{\bullet i}&=&
{gs\over 8\pi^2}
\!\int\! dz_{\perp}{(x-z)^k\over -x_{\parallel}^2+(x-z)_{\perp}^2+i\epsilon}
\left({g_{ik}[Y_j,V^j]\over x_{\ast}-i\epsilon} +
{[Y_i,V_k]-[Y_k,V_i]\over x_{\ast}+i\epsilon}\right)
\nonumber\\
F^{(1)}_{\ast i}&=&-
{gs\over 8\pi^2}
\!\int\! dz_{\perp}{(x-z)^k\over -x_{\parallel}^2+(x-z)_{\perp}^2+i\epsilon}
\left({g_{ik}[Y_j,V^j]\over x_{\bullet}-i\epsilon}-
{[Y_i,V_k]-[Y_k,V_i]\over x_{\bullet}+i\epsilon}\right)\nonumber
\end{eqnarray}

\begin{figure}[htb]
\hspace{1cm}
\mbox{
\epsfxsize=14cm
\epsfysize=4cm
\hspace{0cm}
\epsffile{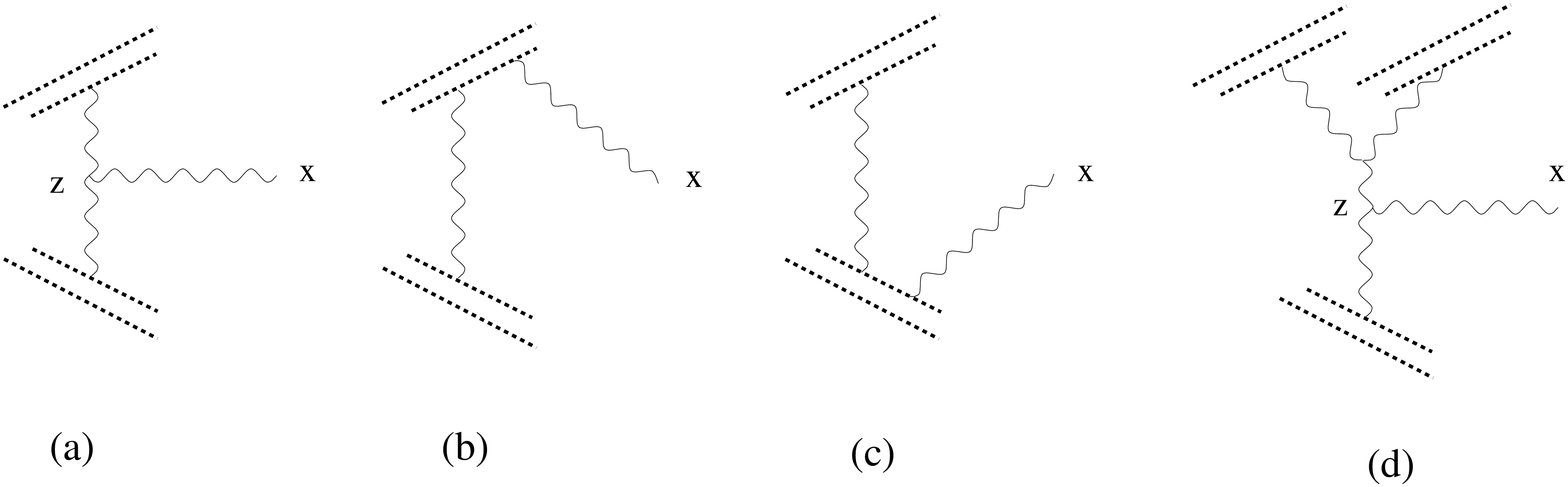}}
\vspace{0cm}
{\caption{Perturbative Feynman diagrams for the field strength
(\ref{fla52})  \label{fig11}}}
\end{figure}

In terms of usual Feynman diagrams (when we expand in powers of source 
just like in previous Section) these expressions come from the diagrams
shown in Fig. \ref{fig11}.
When we sum up the three contributions
from the diagrams in Fig. \ref{fig11}a,b, and c the three-gluon
vertex in Fig. \ref{fig11}a is replaced by the effective Lipatov's vertex 
and we get (\ref{fla52}) up to the terms 
${1\over \partial^2}\partial_i\partial_kY^k$ and 
${1\over \partial^2}\partial_j\partial_kV^k$ standing in place of $Y_i$ 
and $V_j$.
However, as we have discussed in Sect. 3, the difference 
$Y_i-{1\over \partial^2}\partial_i\partial_kY^k=
g{\partial_k\over \partial^2}[Y_i,Y_k]$ (which has an additional power of g) 
will be canceled by the next-order perturbative diagrams 
of the Fig. \ref{fig11}d type.

 Let us now find the effective action. In the trivial order the only non-zero 
field strength components are 
$F^{(0)}_{\bullet i}=\delta({2\over s}x_{\ast})Y_i(x_{\perp})$ and
$F^{(0)}_{\ast i}=\delta({2\over s}x_{\bullet})V_i(x_{\perp})$ 
so we get the familiar  expression 
$S^{(0)}=\int d^2x_{\perp}V^{ai}Y^{a}_i$. In the next order one has
\begin{eqnarray}
&&S^{(1)}=\int d^4x\left(-{2\over s}F_{\ast}^{(1)ai}F^{(1)a}_{\bullet i}-
{1\over 4}F^{(1)a}_{ik}F^{(1)aik}+{2\over s^2}F^{(1)a}_{\ast\bullet}
F^{(1)a}_{\ast\bullet}\right)+2\!\int\! d^2x_{\perp} \nonumber\\
&&
\int\! du\Big({\rm Tr}V^{i}
\left([-\infty p_1,up_1]_{x_{\perp}}
F_{\bullet i}(up_1+x_{\perp})
[up_1+x_{\perp},-\infty p_1]_{x_{\perp}}\right)^{(1)}+\nonumber\\
&&
{\rm Tr}Y^{i} \left(-\infty p_2,up_2]_{x_{\perp}}
F_{\ast i}(up_2+x_{\perp})[up_2,\infty
p_2]_{x_{\perp}}\right)^{(1)}\Big) \label{fla53}
\end{eqnarray} 
We have seen above that the effective action
contains $\ln{\sigma\over\sigma'}$ (see Eq. (\ref{fla23})). 
With logarithmic accuracy, the r.h.s of Eq. (\ref{fla53}) reduces to
\begin{eqnarray}
S^{(1)}&=&-{2\over s}\int d^4x F^{(1)ai}_{\ast}(x)F_{\bullet i}^{(1)a}(x)
\nonumber\\
&+&\int d^2x_{\perp}2{\rm Tr}[Y^{i},V_{i}]
\left([x_{\perp},-\infty p_2+x_{\perp}]^{(1)}-
[x_{\perp},-\infty p_1+x_{\perp}]^{(1)}\right)
\label{fla54}
\end{eqnarray}
The first term contains the integral over
$d^4x={2\over s}dx_{\bullet}dx_{\ast}d^2x_{\perp}$. In order to separate the 
longitudinal divergencies 
from the infrared divergencies in the transverse space we will
work in the $d=2+2\epsilon$ transverse dimensions. 
It is convenient to perform at first the 
integral over $x_{\ast}$ which is determined by a residue in the
point $x_{\ast}=0$. The integration over remaining 
light-cone variable 
$x_{\bullet}$ factorizes then in the form 
$\int_{0}^{\infty}dx_{\bullet}/x_{\bullet}$ or
$\int_{-\infty}^{0}dx_{\bullet}/x_{\bullet}$.
This integral reflects our usual longitudinal logarithmic divergencies
which arise from the replacement of vectors $n$ and $n'$ in (\ref{fla22}) 
by the light-like vectors $p_1$ and $p_2$. 
In the momentum space this logarithmical divergency has the form 
$\int d\alpha/\alpha$. 
It is clear that when $\alpha$ is close to $\sigma$ (or $\sigma'$) we 
can no longer approximate $n$ by $p_1$ (or $n'$ by $p_2$). Therefore, 
in the leading log approximation this divergency should be replaced by 
$\ln{\sigma\over\sigma'}$:
\begin{eqnarray}
\int_{0}^{\infty}dx_{\bullet}{1\over x_{\bullet}}=
\int_{0}^{\infty}d\alpha {1\over \alpha}\rightarrow 
\int_{\sigma}^{\sigma'}d\alpha {1\over \alpha}~=~\ln{\sigma\over\sigma'}
\label{fla55}
\end{eqnarray}
The (first-order) gauge links in the second term in r.h.s. 
of Eq. (\ref{fla54}) have the logarithmic divergence of the same origin:
\begin{eqnarray}
&&[x_{\perp},-\infty p_1+x_{\perp}]^{(1)}=-{i\over 8\pi^2}
\int^{0}_{-\infty}\!{dx_{\ast}\over x_{\ast}}\int d^2x_{\perp}
{\Gamma(\epsilon)\over(x-z)_{\perp}^{2\epsilon}}
[Y_k(z_{\perp}),V^k(z_{\perp})]\nonumber\\
&&[x_{\perp},-\infty p_2+x_{\perp}]^{(1)}={i\over 8\pi^2}
\int^{0}_{-\infty}\!{dx_{\bullet}\over x_{\bullet}}\int d^2x_{\perp}
{\Gamma(\epsilon)\over(x-z)_{\perp}^{2\epsilon}}
[Y_k(z_{\perp}),V^k(z_{\perp})]
\label{fla56}
\end{eqnarray}
which also should be replaced by $\ln{\sigma\over\sigma'}$.
Performing the remaining integration over $x_{\perp}$ in the first term
in r.h.s. of Eq. (\ref{fla54}) we obtain the 
the first-order classical action in the form:
\begin{eqnarray}
\lefteqn{S^{(1)}=}\label{fla57} \\
&-{ig^2\over 8\pi^2}\ln{\sigma\over\sigma'}
\int d^2x_{\perp}
d^2y_{\perp}\big(L^a_1(x_{\perp})L^a_1(y_{\perp})+
L^a_2(x_{\perp})L^a_2(y_{\perp})\big)
{\Gamma(\epsilon)\over(x-y)_{\perp}^{2\epsilon}}
\nonumber
\end{eqnarray}
where
\begin{eqnarray}
L^a_1&=&if^{abc}Y^a_jV^{bj}, L^a_2=i\epsilon_{ik}Y^{ai}V^{bk}
\label{fla58}
\end{eqnarray}
and $\epsilon_{ik}$ is the totally antisymmetric tensor in two transverse
dimensions ($\epsilon_{12}=1$). One may also rewrite this 
expression in a compact form
\begin{eqnarray}
S^{(1)}&=&{ig^2\over 2\pi}\ln{\sigma\over\sigma'}
\int d^2x_{\perp} \left(L_1^a{1\over \partial_{\perp}^2}L_1^a+
L_2^a{1\over \partial_{\perp}^2}L_2^a\right)
\label{fla59}
\end{eqnarray}

A more accurate version of this formula looks like:
\begin{eqnarray}
S^{(1)}&=&\nonumber\\
&~&{ig^2\over 2\pi}\ln{\sigma\over\sigma'}
\int d^2x_{\perp}
\Big(L_1^a{1\over \partial_{\perp}^2}L_1^a+
L_2^a\big(Y^{\dagger}{1\over \partial_{\perp}^2}Y+
V^{\dagger}{1\over \partial_{\perp}^2}V-
{1\over \partial_{\perp}^2}\big)^{ab}L_2^b+
\nonumber\\
&~&
L_1^a\big({\partial_i\over\partial^2}Y^{\dagger}{\partial_k\over\partial^2}Y
-Y\leftrightarrow V\big)L^b_2\epsilon^{ik} -L_2^a\epsilon^{ik}
\big(Y^{\dagger}{\partial_i\over\partial^2}Y{\partial_k\over\partial^2}
-Y\leftrightarrow V\big)^{ab}L^b_1\Big)+\nonumber\\
&~&O([U,V]^3)\label{fla60}
\end{eqnarray}
where
\begin{eqnarray}
Y(x_{\perp})=[\infty p_1,-\infty p_1]_{x_{\perp}},~~~~~~~
V(x_{\perp})=[\infty p_2,-\infty p_2]_{x_{\perp}}\label{fla61}
\end{eqnarray}
 It is easy to see that in the case of one weak and one strong  
 source this expressions coincides with (\ref{dob5}) (up to the terms of higher 
 order in weak source which we neglect anyway).
 
\begin{figure}[htb]
\hspace{4cm}
\mbox{
\epsfxsize=6cm
\epsfysize=4cm
\hspace{0cm}
\epsffile{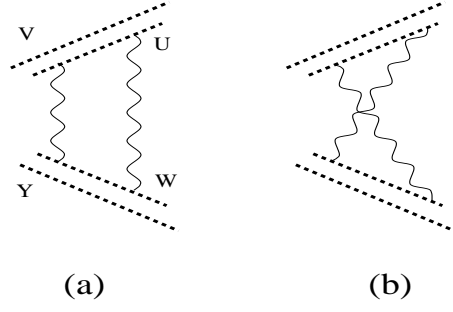}}
\vspace{1cm}
{\caption{Lowest-order diagrams for gluon reggeization. \label{figreg}}}
\end{figure}

At $d=2$ we have an infrared pole in $S^{(1)}$ which must be canceled by the 
corresponding divergency in the trajectory of the reggeized gluon. 
The gluon reggeization is not a classical effect in our approach - rather, 
it is a quantum correction coming from the loop corresponding to the determinant
of the operator of second derivative of the action
\begin{equation}
\left.{\delta \over \delta A_{\mu}}{\delta \over \delta
A_{\nu}}\left(S_{QCD}+\!\int\! d^2x_{\perp} 
V^{ai}(x_{\perp})U^{a}_i(x_{\perp}) +\!\int\!
d^2x_{\perp}W^{ai}Y^a_i(x_{\perp}\right)\right|_{A=\bar{A}} \label{fla62}
\end{equation} 
The lowest-order diagrams are shown in 
Fig. \ref{figreg}
and the explicit form of the second derivative 
of the Wilson-line operator is:
\begin{eqnarray}
&\delta U_i=i
\int^{\infty}_{-\infty}du\int^{u}_{-\infty}dv
[\delta\hat{A}_i(u p_1+x_{\perp}),
\hat{\nabla}_i\delta\hat{A}_i(v p_1+x_{\perp})]\nonumber\\
&\delta W_i=i
\int^{\infty}_{-\infty}du\int^{u}_{-\infty}dv
[\tilde{A}_i(u p_2+x_{\perp}),
\tilde{\nabla}_i\delta\tilde{A}_i(u p_2+x_{\perp})]
\label{fla63}
\end{eqnarray}
Now one easily gets the contribution of the 
Fig. \ref{figreg} diagrams in the form:
\begin{eqnarray}
\lefteqn{S_{\rm r}=}\label{fla64}\\
&{g^2N_c\over 8\pi^3}\ln{\sigma\over\sigma'}
\int d^2x_{\perp} d^2y_{\perp} \big(V_i^a(x_{\perp})Y^{ai}(y_{\perp})-
V_i^a(x_{\perp})Y^{ai}(x_{\perp})\big)
{\Gamma^2(1+\epsilon)\over((x-y)_{\perp}^2)^{(1+2\epsilon)}}
\nonumber
\end{eqnarray}
A more accurate form
of this equation reads:
\begin{eqnarray}
\lefteqn{S_{\rm r}=}\label{fla65}\\
&-&{g^2N_c\over 8\pi^3}\ln{\sigma\over\sigma'}
\int d^2x_{\perp} d^2y_{\perp} 
{\Gamma^2(1+\epsilon)\over((x-y)_{\perp}^2)^{(1+2\epsilon)}}
\Big\{V_i^a(x_{\perp})Y^{ai}(x_{\perp})-\nonumber\\
&~&
{1\over N_c}\Big( Y^{i}(x_{\perp})\{Y(x_{\perp})Y^{\dagger}(y_{\perp})+
V(x_{\perp})V^{\dagger}(y_{\perp})-1\}
Y^{i}(y_{\perp})\Big)^{aa}\Big\}+O([U,V]^3)\nonumber
\end{eqnarray}
where ${\cal O}^{aa}\equiv$Tr$O$ in the gluonic representation. In the 
case of one strong and one weak source it coincides 
with (\ref{dob2}) (up
to the higher powers of weak source).
 
The complete first-order ($\equiv$ one-log) expression 
for the effective action is
the sum of $S^{(0)}$, $S^{(1)}$, and $S_{\rm r}$:
\begin{eqnarray}
\lefteqn{S_{\rm eff}=
\int d^2xV^{ai}(x)Y^{a}_i(x)-}\label{fla66}\\
&&{ig^2\over 8\pi^2}\ln{\sigma\over\sigma'}\!\int\! d^2x
d^2y\Big\{{\Gamma(\epsilon)\over(x-z)^{2\epsilon}}
\Big(L^a_1(x)L^{a}_1(y)+L^a_2(x)L^{b}_2(y)
\big(Y^{\dagger}_xY_y +V^{\dagger}_xV_y-1\big)^{ab}\Big)
\nonumber\\ 
&&\int\! d^2z 
{\epsilon^{ij}(x-z)_i(z-y)_j\over \pi^2(x-z)^2(z-y)^2}
\Big(L^a_1(x)\big(Y^{\dagger}_zY_y -Y\leftrightarrow V\big)^{ab}L^{b}_2(y)-
\nonumber\\ 
&&
L^a_2(x)\big(Y^{\dagger}_xY_z -Y\leftrightarrow V\big)^{ab}L^{b}_1(y)\Big)
\Big\}\nonumber\\ 
&&-{g^2N_c\over 8\pi^3}\ln{\sigma\over\sigma'}
\int d^2x_{\perp} d^2y_{\perp} 
{\Gamma^2(1+\epsilon)\over((x-y)_{\perp}^2)^{(1+2\epsilon)}}
\Big\{V_i^a(x_{\perp})Y^{ai}(x_{\perp})-\nonumber\\
&&
{1\over N_c}\Big( V^{i}(x_{\perp})\{Y(x_{\perp})Y^{\dagger}(y_{\perp})+
V(x_{\perp})V^{\dagger}(y_{\perp})-1\}
Y^{i}(y_{\perp})\Big)^{aa}\Big\}+O([U,V]^3)\nonumber
\end{eqnarray}
At one weak and one large source it coincides with (\ref{dob6}). (As we
discussed in Sect. 4, the
new nontrivial terms in the case of two strong sources start from
$[U,V]^3\ln^2{\sigma\over\sigma'}$).  

As usual, in the case of scattering of white objects the logarithmic
infrared divergence $\sim {1\over\epsilon}$ cancels. 
For example, for the case of one-pomeron exchange the relevant term
in the expansion of $e^{iS_{\rm eff}}$ has the form:
\begin{eqnarray}
&-{g^2\over 16\pi^2}\ln{\sigma\over\sigma'}\int d^2x_{\perp} d^2y_{\perp} 
f^{dam}(V_j^aY^{mj}g_{ik}+V^a_iY^m_k-V^a_kY^m_i)(x_{\perp})\nonumber\\
&
{\Gamma(\epsilon)\over(x-y)_{\perp}^{2\epsilon}}
f^{dbn}(V_l^bY^{nl}g^{ik}+V^{bi}Y^{mk}-V^{bk}Y^{mi})(y_{\perp})+\nonumber\\
&
{g^2N_c\over 16\pi^3}\ln{\sigma\over\sigma'}
\int d^2x_{\perp}V^a_i(x_{\perp})Y^{ai}(x_{\perp})\int d^2y_{\perp} 
d^2y'_{\perp} (V_j^b(y_{\perp})-V_j^b(y'_{\perp}))
\nonumber\\
&{\Gamma^2(1+\epsilon)\over((y-y')_{\perp}^2)^{(1+2\epsilon)}}
(Y^{bj}(y_{\perp})-Y^{bj}(y'_{\perp}))
\label{fla67}
\end{eqnarray}
It is easy to see that the terms $\sim {1\over\epsilon}$  cancel if 
we project onto 
colorless state in t-channel (that is, replace $V^{ai}V_j^b$ by 
${\delta_{ab}\over N_c^2-1}V^{ci}V_j^c$). It is worth noting that in the two-gluon
approximation the r.h.s. of the eq. (\ref{fla67}) gives the BFKL kernel.

\section{Conclusions and outlook} 

The ultimate goal of this approach is to obtain the explicit expression
for the effective action in all orders in $ln{s\over m^2}$. One possible
prospect is that due to the conformal invariance of QCD at the tree level our 
future result for the effective action can be formalized in terms of conformal 
two-dimensional theory in external two-dimentional ``gauge fields'' 
$V_i$ and $Y_i$. 

Up to now, we have not used the conformal invariance because
it is not obvious how to implement it in terms of Wilson-line operators.
We can, however, expand Wilson lines 
back to gluons. The conformal properties of (reggeized) gluon
amplitudes are well studied now. In the coordinate space the BFKL kernel
is invariant under Mobius group and therefore the eigenfunctions of 
BFKL kernel are simply powers of coordinates. Moreover, at large $N_c$ the 
diagrams with 
fixed number of
 reggeized gluons (which form a {\it unitary} subset of all diagrams) may
 be described in terms of two-dimensional quantum mechanics of the particles 
with Lipatov's
 Hamiltonian (\ref{fla23}). Due to the property of a holomorphic separability
 this two-dimensional quantum
 mechanics reduces to the one-dimesional Heisenberg xxx spin-0 model
 \cite{lkf}. 
(Unfortiunately, 
the exact solution of this model is not known yet). It is not clear which part
of this symmetry survives for the  full effective action but there is every
 reason to believe that it will simplify 
the structure of the answer even after reassembling of Wilson lines.

In conclusion I would like to note that the semiclassical approach 
developed above for the small-x processes in perturbative QCD
 may 
be modified for studying the heavy-ion collisions. As advocated in 
ref. \cite{larry2}, for the
heavy-ion collisions the coupling constant 
may be relatively small due to high density. 
On the other hand, the fields produced by colliding ions are large so 
that the product $gA$ is not small -- 
which means that the Wilson-line gauge factors $V$ and $Y$ are of order of 1. 
It should be mentioned, however, that in this paper 
we considered the special case 
of the collision of the two shock waves, namely without any particles 
in the final state. It follows from the usual boundary conditions for Feynman 
amplitude (\ref{fla6}) which we calculate: 
no outgoing waves at $t\rightarrow\infty$ (and no incoming fields at 
$t\rightarrow-\infty$,
but we have satisfied this condition by choosing the gauge 
$\left.A\right|_{t\rightarrow-\infty}=0$). 
However, people are usually interested in the process of particle production  
during the collision (see e.g. \cite{almuller}) since it gives the experimental 
probe of quark-gluon plasma. 
In this case, our approach must be modified for the new boundary conditions
---
we must solve the  classical equations (\ref{fla48}) with only half of the 
boundary conditions (\ref{fla49}) at $t\rightarrow-\infty$. 
The boundary condition
at $t\rightarrow\infty$ depends on the problem under investigation: in the 
case if we are interested in the 
wavefuction of the system at large times we do not have any boundary conditions
at $t\rightarrow\infty$ but we must use the causal (retarded and advanced) Green 
function instead of the usual Feynman ones. ( For example, 
in the expression (\ref{fla52}) for the field strength we will 
have the retarded Green function 
$\theta(x_0)2\pi\delta(x_{\parallel}^2-(x-z)_{\perp}^2)$ 
instead
of the Feynman propagator $(-x_{\parallel}^2+(x-z)_{\perp}^2+i\epsilon)^{-1}$) 
On the contrary, if we calculate 
the total cross section (cut diagrams) we must calculate the 
double functional integral 
corresponding to the integration over the ``+'' fields to the right 
and the ``-'' fields to the left of the cut (see ref\cite{keld}). 
(This is actually a functional-integral 
formalization of Cutkovsky rules).
In this case we may use the usual (Feynman and c.c. Feynman)
propagators for each type of the fields. 
The boundary condition 
 requires that two
types of the field --- the left -side ``-'' fields and the right-side ``+''
ones  --- coincide at $t\rightarrow\infty$.  (This boundary condition is
responsible for the $\delta(p^2)\theta(p_0)$  propagators on the cut). 
Thus, to find the total cross section of the shock-wave collision in the 
semiclassical 
approximation we must solve 
the double set of classical equations for ``+'' and ``-'' fields with the 
boundary condition that these fields coincide at infinity. The study is in progress.

\vskip0.5cm
\section*{Acknowledgments}

The author is grateful to Y. Kovchegov, L.N. Lipatov, L. McLerran 
and A.V. Radyushkin for valuable discussions.
 
This work was supported by the US Department of Energy under contract 
DE-AC05-84ER40150.


\section*{References}
\begin{thebibliography}{99}

\bibitem{bfkl}
  V.S. Fadin, E.A. Kuraev, and L.N. Lipatov, \Journal{\PLB}{60}{50}{1975};\\
 I.I. Balitsky and L.N. Lipatov, \Journal{\em Sov. Journ. Nucl. Phys.} 
{28}{822}{1978}

\bibitem{nlobfkl}
V.S. Fadin and L.N. Lipatov,\Journal{\PLB}{429}{127}{1998};
\bibitem{ing}
  I. Balitsky, \Journal{\NPB}{463}{99}{1996}.

\bibitem{verlinde}
   H. Verlinde and E. Verlinde,
   ``QCD at High Energies and Two-Dimensional Field Theory'',
   preprint PUPT-1319, e-Print Archive: hep-th/9302104.

\bibitem{fak}
  J.C. Collins, D.R. Soper, and G. Sterman,
"Factorization of Hard Processes in QCD",
in {\it Perturbative QCD}, ed. A.H. Mueller (World Scientific, 
Singapore, 1989) 

\bibitem{ing}
  I. Balitsky, 
  \Journal{\NPB}{463}{99}{1996}.

\bibitem{lobzor}
  L.N. Lipatov, 
\Journal{\em Phys. Reports} {286}{131}{1997}.

\bibitem{eveq}
  I. Balitsky abnd V.M. Braun, \Journal{\NPB}{311}{541}{1989}. 

\bibitem{prl}
  I. Balitsky, \Journal{\PRL}{81}{2024}{1998}.

\bibitem{larry1}
L. McLerran and R. Venugopalan, 
\Journal{\PRD}{50}{2225}{1994};  
A. Ayala, J. Jalilian-Marian, L. McLerran , and
R. Venugopalan, \Journal{\PRD}{52}{2935}{1995}.

\bibitem{nacht}
O. Nachtmann, \Journal{\em Ann. Phys.} {209}{436}{1991}.


\bibitem{mes}
  I.I. Balitsky and L.N. Lipatov, 
\Journal{\em JETP Letters} {30}{355}{1979}.

 
\bibitem{wlup}
  I.I. Balitsky, \Journal{\NPB}{254}{166}{1985}.

\bibitem{dosch}
  H.G. Dosch, E. Ferreira, and A. Kraemer,
\Journal{\PRD}{50}{2015}{1994}.

\bibitem{tok}
  I. Balitsky,
"Factorization and Effective action for High-Energy Scattering".\\
To be published in the Proceedings of the 3rd Workshop on 
Continuous Advances in QCD (QCD 98), Minneapolis; 
e-print archive: hep-ph/9808215
 

\bibitem{kovner}
A. Kovner, L. McLerran and H. Weigert, 
\Journal{\PRD}{52}{6231}{1995}.


\bibitem{many}
  R. Kirschner, L.N. Lipatov, L. Szymanowski, 
\Journal{\NPB}{425}{579}{1994};
     L.N. Lipatov, \Journal{\NPB}{452}{369}{1996}
  
\bibitem{l}
  L.N. Lipatov, \Journal{\em Sov. Phys. JETP} {63}{904}{1986}. 

\bibitem{lkf}
  L.N. Lipatov, \Journal{\em JETP Letters} {59}{571}{1994};
  L.D. Faddeev and G.P. Korchemsky,
  \Journal{\PLB}{342}{311}{1995}.


\bibitem{larry2}
L. McLerran and R. Venugopalan, \Journal{\PRD}{49}{2233}{1994}
\Journal{\PRD}{49}{3352}{1994}.

\bibitem{almuller}
 Yu.V. Kovchegov, A.H. Mueller, \Journal{\NPB}{529}{451}{1998}. 

\bibitem{keld}
 I. Balitsky and V.M. Braun, \Journal{\NPB}{361}{93}{1991}, 
 \Journal{\NPB}{380}{51}{1992}
\end{thebibliography}
\end{document}